\documentclass[12pt]{article}

\usepackage{psfrag,epsf}
\usepackage{enumerate}
\usepackage[slantedGreek]{mathpazo}
\usepackage[pdftex]{graphicx}
\usepackage{amsfonts}
\usepackage{setspace}
\usepackage{amssymb}
\usepackage{amsthm}
\usepackage{amsmath}
\usepackage{mathtools}
\usepackage{mathrsfs}
\usepackage{caption}
\usepackage{subcaption}
\usepackage{color}
\usepackage{hyperref}
\usepackage{syntonly}
\usepackage{float}
\usepackage{booktabs}
\usepackage{bm}
\usepackage{algorithm}
\usepackage{algpseudocode}
\usepackage{bbold}
\usepackage[export]{adjustbox}
\usepackage{setspace}
\usepackage{titlesec}
\usepackage{lipsum}
\usepackage{natbib}
\usepackage{tabularx}
\usepackage{graphicx}
\usepackage{mwe}
\allowdisplaybreaks

\usepackage[utf8]{inputenc}
\usepackage[english]{babel}

\newtheorem{theorem}{Theorem}[section]

\newtheorem{lemma}[theorem]{Lemma}

\newcommand\blfootnote[1]{%
  \begingroup
  \renewcommand\thefootnote{}\footnote{#1}%
  \addtocounter{footnote}{-1}%
  \endgroup
}

\newcommand{\blind}{0}

\begin{document}

\def\spacingset#1{\renewcommand{\baselinestretch}%
{#1}\small\normalsize} \spacingset{1}

\if0\blind
{
  \title{\bf The Graphical Horseshoe Estimator for Inverse Covariance Matrices}
  \author{Yunfan Li \\
    Department of Statistics, Purdue University \\
    and \\
    Bruce A. Craig \\
    Department of Statistics, Purdue University \\
    and \\
    Anindya Bhadra \\
    Department of Statistics, Purdue University
    \blfootnote{Yunfan Li (e-mail: li896@purdue.edu) is Ph.D. Candidate, Department of Statistics, Purdue University; Bruce A. Craig is Professor, Department of Statistics, Purdue University; and Anindya Bhadra is Associate Professor, Department of Statistics, Purdue University, West Lafayette, IN 47907, USA.}
    }
  \maketitle
} \fi

\if1\blind
{
  \bigskip
  \bigskip
  \bigskip
  \begin{center}
    {\LARGE\bf The Graphical Horseshoe Estimator for Inverse Covariance Matrices}
\end{center}
  \medskip
} \fi

\bigskip
\begin{abstract}
We develop a new estimator of the inverse covariance matrix for high-dimensional multivariate normal data using the horseshoe prior. The proposed graphical horseshoe estimator has attractive properties compared to other popular estimators, such as the graphical lasso and the graphical smoothly clipped absolute deviation. The most prominent benefit is that when the true inverse covariance matrix is sparse, the graphical horseshoe provides estimates with small information divergence from the sampling model. The posterior mean under the graphical horseshoe prior can also be almost unbiased under certain conditions. In addition to these theoretical results, we also provide a full Gibbs sampler for implementing our estimator. MATLAB code is available for download from github at \textit{http://github.com/liyf1988/GHS}. The graphical horseshoe estimator compares favorably to existing techniques in simulations and in a human gene network data analysis.
\end{abstract}

\noindent%
{\it Keywords:}  Bayesian shrinkage estimation; Gaussian graphical model; high-dimensional graphs; sparse precision matrix.
\vfill


\newpage

\spacingset{1.45}

\section{Introduction}
\label{sec:intro}
		
Estimation of the covariance or inverse covariance matrix of a multivariate normal vector plays a central role in numerous fields, including spatial data analysis \citep{cressie2015statistics}, variance components and longitudinal data analysis \citep{diggle2002analysis}, and the growing area of genetic data analysis \citep{dehmer2008analysis}. \citet{pourahmadi2011covariance} provides a survey of some of the most popular methods in high-dimensional covariance and inverse covariance estimation. In a penalized likelihood framework, two of the most notable methods for inverse covariance estimation are the graphical lasso \citep{friedman2008glasso} and the graphical SCAD \citep{fan2009network}. Both these methods provide estimates for a high-dimensional inverse covariance matrix under an arbitrary sparsity pattern.

There has also been much recent work in covariance and inverse covariance estimation in a Bayesian framework. \citet{banerjee2014posterior} proposed a prior distribution for estimating a banded inverse covariance matrix. \citet{rajaratnam2008flexible} and \citet{xiang2015high} proposed Bayesian estimators for the covariance of a decomposable Gaussian graphical model. \citet{pati2014posterior} considered sparse factor models for covariance matrices and induced a class of continuous shrinkage priors on the factor loadings. There are also studies that focus on the theoretical properties of these estimators, including posterior convergence rates, Bayesian minimax rates and consistency of Bayesian estimators \citep{banerjee2014posterior, banerjee2015bayesian, xiang2015high, lee2017estimating, lee2017optimal}. However, to our knowledge, few Bayesian estimators assume an arbitrary sparsity pattern of the true inverse covariance matrix. Under such an assumption, \citet{banerjee2015bayesian} proposed a mixture prior for graphical structure learning, and \citet{wang2012bayesiangl} developed a Bayesian version of the graphical lasso.

In this paper, we propose an alternative Bayesian estimator, which we call the graphical horseshoe estimator. This estimator works under the assumption of an arbitrary sparsity pattern in the inverse covariance matrix. We show that our estimator has better performance in adapting to sparsity in high-dimensional problems than some competing methods because of two properties of our prior: greater concentration near the origin and heavier tails. Both of these properties are inherited from the horseshoe prior of \citet{carvalho2010horseshoe} for the sparse normal means model.

Many attractive theoretical properties of the horseshoe prior have been discovered in recent years for the normal means model. These include improved Kullback--Leibler risk bounds \citep{carvalho2010horseshoe}, asymptotic optimality in testing under $0-1$ loss \citep{datta2013asymptotic}, minimaxity in estimation under the $\ell_2$ loss \citep{van2014horseshoe}, and improved risk properties in linear regression \citep{bhadra2016prediction}. In this paper, we demonstrate how some of these properties translate to the estimation of the inverse covariance matrix in a multivariate Gaussian model. We discuss the implications of these properties both theoretically and empirically.

The remainder of this paper is organized as follows. The rest of Section~\ref{sec:intro} discusses three competing methods for sparse precision matrix estimation: the graphical lasso, the graphical SCAD, and the Bayesian graphical lasso. Section~\ref{sec:GHS} outlines the graphical horseshoe estimator as well as a full Gibbs sampler for easy and efficient sampling. Sections~\ref{sec:KL} and \ref{sec:bias} outline the theoretical properties of our proposed estimator along with a comparison to the graphical lasso and graphical SCAD estimators. Section~\ref{sec:sim} illustrates these theoretical properties through simulations. Section~\ref{sec:real data} applies the proposed method on a human gene expression data set to identify a sparse gene interaction network, before concluding with some discussion of possible future research topics in Section~\ref{sec:conc}.
	
\subsection{Related Works in Precision Matrix Estimation}
\label{sec:model}	
Consider $n$ samples from a $p$-dimensional multivariate normal distribution with zero mean and a $p \times p$ covariance matrix $\Omega^{-1}$. That is,
\begin{align*}
\mathbf{y}_k \sim \mathrm{Normal}(\mathbf{0},\Omega^{-1}),
\end{align*}
for $k=1,\ldots,n$. Under this parameterization, the inverse of the covariance matrix, $\Omega$, is referred to as the precision matrix (assumed to be positive definite). The $ij$th off-diagonal element in $\Omega$ is the negative of the partial covariance between features $i$ and $j$, and the $i$th diagonal element is the inverse of the residual variance when the $i$th feature is regressed on all the other features \citep{pourahmadi2011covariance}.  Under the multivariate normal model, zero off-diagonal elements in $\Omega$ correspond to features that are conditionally independent given the remaining features. In certain applications, estimating the precision matrix is attractive, especially in high-dimensional cases, since it is preferable to study partial correlations rather than marginal correlations \citep{pineda2014guiding}.
	
A major challenge in precision matrix estimation is that the number of free parameters grows quadratically with the number of features. As a consequence, in high-dimensional problems, some methods assume the covariance or precision matrix has a structure, such as latent factors \citep{pati2014posterior} or banding \citep{banerjee2014posterior}. When the structure of the true precision matrix is assumed to be arbitrary, the precision matrix is usually assumed to be sparse. In high-dimensional settings, a natural approach for estimating a sparse model is to penalize the likelihood. \citet{friedman2008glasso} proposed the graphical lasso, which estimates the precision matrix under the lasso penalization \citep{tibshirani1996lasso} while maintaining the symmetry of the estimate. The graphical lasso maximizes the penalized likelihood:
\begin{equation}  \label{eq:1}
	\log(\det\Omega)-\mathrm{tr}(S\Omega/n)-\sum_{i,j} \phi_{\lambda}(|\omega_{ij}|),
\end{equation}	
where $S=\sum_{i=1}^n \mathbf{y}_i\mathbf{y}_i‘$ is the scatter matrix, $\Omega=(\omega_{ij})$, $\phi_{\lambda}(|\omega_{ij}|)=\lambda|\omega_{ij}|$ is the $\ell_1$ penalty, and $\lambda$ is a tuning parameter. In practice, $\lambda$ is often chosen by cross validation. The sum $\sum_{i,j} \phi_{\lambda}(|\omega_{ij}|)$ in Equation~(\ref{eq:1}) can be taken with or without a penalty on the diagonal terms \citep{rothman2008sparse, meinshausen2006mh, yuan2007model, friedman2008glasso}.

A Bayesian version of graphical lasso was proposed by \citet{wang2012bayesiangl}. In the Bayesian setting, the frequentist graphical lasso estimator is equivalent to the maximum a posteriori estimate of $\Omega$ under the following prior:
\begin{equation*}
	p(\Omega \, | \, \lambda) \propto \prod_{i<j}\{\mathrm{DE}(\omega_{ij} \, | \, \lambda)\} \prod_{i=1}^p \{\mathrm{EXP}(\omega_{ii} \, | \, \lambda/2) \} 1_{\Omega \in \mathcal{S}_p}\;,
\end{equation*}
where $\mathrm{DE}(x \, | \, \lambda)$ represents the double exponential distribution with rate $\lambda$, $\mathrm{EXP}(x \, | \, \lambda)$ represents the exponential distribution with rate $\lambda$, and $\mathcal{S}_p$ is the space of $p \times p$ positive definite matrices. The tuning parameter $\lambda$, or rather the hyper-parameter in the language of Bayesian hierarchical models, can be chosen by cross-validation as in a frequentist framework \citep{friedman2008glasso, rothman2008sparse}, or by a fully Bayesian method with an appropriate hyperprior.
	
The smoothly clipped absolute deviation (SCAD) penalty by \citet{fan2001scad} was introduced in precision matrix estimation because of its attractive asymptotic properties. The graphical SCAD maximizes the penalized likelihood in Equation~(\ref{eq:1}) where the penalty has the first order derivative:
	\begin{equation*}
		\phi'_{\lambda}(|x|) =
		\lambda \left\{ 1_{\{|x|\leq \lambda \}} + \frac{(a\lambda-|x|)_+}{(a-1)\lambda}1_{\{|x|>\lambda\}} \right\},
	\end{equation*}
with $a>2$ and $\lambda>0$. This penalty is linear near the origin and non-decreasing. In practice, the tuning parameter $a$ is often fixed while $\lambda$ is chosen by cross validation. The graphical SCAD estimate satisfies the oracle property given by \citet{fan2001scad}. The SCAD penalty does not have a Bayesian representation, although \citet{polson2012levy} provide an understanding of how priors and penalty functions are related even when some penalty functions lack Bayesian equivalents. \citet{lam2009sparsistency} showed that under certain conditions, both frequentist graphical lasso and graphical SCAD estimates of the precision matrix converge to the true precision matrix under the Frobenius norm. However, these theoretical results depend on theoretical choices of tuning parameters, which cannot be implemented in practice. The regulatory conditions are also difficult to check in data analysis.
	
All methods for large sparse precision matrix estimation face the problem of accumulated estimation errors due to the large number of parameters to estimate. Furthermore, the double-exponential priors in the Bayesian lasso have been shown to possess some undesirable properties in the high-dimensional normal means problem \citep{carvalho2009handling, carvalho2010horseshoe}. Although lasso and SCAD are widely-used methods with good asymptotic properties, the element-wise bias of graphical lasso estimates can be large, and graphical SCAD does not guarantee positive definite estimates \citep{fan2016overview}.

To provide an alternative that remedies the accumulation of errors in high dimensions, we propose a method that obtains a sparse estimate while controlling the element-wise bias of the nonzero elements. In terms of sampling, our method follows the technique adopted in the Bayesian graphical lasso by \citet{wang2012bayesiangl}. However, our method is more efficient at utilizing the sparsity of the precision matrix than the graphical lasso and the graphical SCAD, for reasons we detail in Section~\ref{sec:KL}. Our method also guarantees positive definite and symmetric estimates.

\section{The Graphical Horseshoe Estimator}
\label{sec:GHS}	
		
	Since an unstructured precision matrix is assumed to be sparse, a shrinkage method should be able to give a zero or very small estimate for the zero elements. Meanwhile, a method should also be able to distinguish the non-zero elements in the precision matrix and shrink them as little as possible. We propose the use of the horseshoe prior to do just this.
	
	\subsection{The Graphical Horseshoe Hierarchical Model}
	\label{subsec:model}

	The graphical horseshoe model puts horseshoe priors on the off-diagonal elements of the precision matrix, and an uninformative prior on the diagonal elements, while respecting the constraint $\Omega \in \mathcal{S}_p$. Because the precision matrix is symmetric, we only consider the upper off-diagonal elements. The element-wise priors are specified for $i,j=1,\ldots,p$ as follows:
	\begin{align*}
		\omega_{ii} &\propto 1, \\
		\omega_{ij:i<j} &\sim \text{Normal}(0,\lambda_{ij}^2\tau^2), \\
		\lambda_{ij:i<j} &\sim C^+(0,1), \\
		\tau &\sim C^+(0,1),
	\end{align*}	
	where $C^+(0,1)$ denotes a half-Cauchy random variable with density $p(x) \propto (1+x^2)^{-1}; \ x>0$. The normal scale mixtures with half-Cauchy hyperpriors on the off-diagonal elements is the horseshoe prior proposed by \citet{carvalho2010horseshoe}. The distinctive scale parameter $\lambda_{ij}$ on each dimension is referred to as the local shrinkage parameter, and the scale parameter $\tau$ shared by all dimensions is referred to as the global shrinkage parameter. The marginal prior's peak near the origin induces efficient shrinkage of noise terms in a high-dimensional problem, and the slow decaying tail ensures that signal terms are shrunk very little \citep{carvalho2010horseshoe}.
	
	Thus, the prior on $\Omega$ under graphical horseshoe model can be written as:
	\begin{align*}
		p(\Omega \, | \, \tau) \propto \prod_{i<j} {\mathrm{Normal}(\omega_{ij} \, | \, \lambda_{ij}^2,\tau^2) \prod_{i<j}\mathrm{C}^+(\lambda_{ij} \, | \, 0,1) 1_{\Omega \in \mathcal{S}_p}},
	\end{align*}
	where $\mathcal{S}_p$ is the space of $p \times p$ positive definite matrices. Using the properties of the horseshoe prior \citep{carvalho2010horseshoe}, the induced marginal prior on $\omega_{ij}$ is proper. When $\Omega \in \mathcal{S}_p$, the diagonal elements in $\Omega$ are finite. Therefore the graphical horseshoe prior is proper. In a univariate normal case, the induced marginal prior for shrinkage has infinite mass near both $0$ and $1$ and is thin in between, with a ``horseshoe'' shape \citep{carvalho2010horseshoe}.

	 In high-dimensional precision matrix estimation by the graphical horseshoe, the global shrinkage parameter $\tau$ adapts to the sparsity of the entire matrix $\Omega$ and shrinks the estimates of the off-diagonal elements toward zero. On the other hand, the local shrinkage parameters $\lambda_{ij:i<j}$ preserve the magnitude of non-zero off-diagonal elements, and ensure that the element-wise biases are not very large.
	
	\subsection{A Data-augmented Block Gibbs Sampler}
	\label{subsec:Gibbs}

	Posterior samples under the graphical horseshoe hierarchical model are drawn by an augmented block Gibbs sampler, adapting the scheme proposed by \citet{makalic2016samplerHS} for linear regression. Augmented variables $\nu_{ij: i<j}$ and $\xi$ are introduced for conjugate sampling of the shrinkage parameters $\lambda_{ij:i<j}$ and $\tau$. In each iteration, each column and row of $\Omega$, $\Lambda=(\lambda^2_{ij})$, and $N=(\nu_{ij})$ are partitioned from a $p \times p$ matrix of parameters and updated in a block. Then the global shrinkage parameter $\tau$ and its auxiliary variable $\xi$ are updated.
	
	The following part derives the posterior distribution of the precision matrix. Given data $Y_{n\times p}$ and the shrinkage parameters, the posterior of $\Omega$ under the graphical horseshoe model is
	\begin{align*}
		p(\Omega\,|\, Y,\Lambda,\tau) \propto |\Omega|^{\frac{n}{2}} \text{exp}\Big\{-\text{tr}\Big(\frac{1}{2}S\Omega\Big)\Big\}
		\prod_{i<j}\text{exp}\Big(-\frac{\omega_{ij}^2}{2\lambda_{ij}^2\tau^2}\Big) 1_{\Omega \in \mathcal{S}_p}.
	\end{align*}
	It is not obvious how to sample from this distribution. Following \citet{wang2012bayesiangl}, one column and row of $\Omega$ are updated at a time. Without loss of generality, the posterior distributions for the last column and the last row are derived here. First, partition the last column and row in the matrix:
	\begin{align*}
	   \Omega=
	   \left( {\begin{array}{cc}
	   	\Omega_{(-p)(-p)} & \bm{\omega}_{(-p)p} \\  \bm{\omega}'_{(-p)p} & {\omega}_{pp} \  \end{array} } \right), \
	   S=
	   \left( {\begin{array}{cc}
	    S_{(-p)(-p)} & \mathbf{s}_{(-p)p} \\  \mathbf{s}'_{(-p)p} & {s}_{pp} \      \end{array} } \right), \
	   \Lambda=
	   \left( {\begin{array}{cc}
	   	\Lambda_{(-p)(-p)} & \bm{\lambda}_{(-p)p} \\  \bm{\lambda}'_{(-p)p} & 1 \  \end{array} } \right), \	
	\end{align*}
	where $(-p)$ denotes the set of all indices except for $p$, and $\Lambda_{(-p)(-p)}$ and $\bm{\lambda}_{(-p)p}$ have entries $\lambda_{ij}^2$. Diagonal elements of $\Lambda_{(-p)(-p)}$ can be arbitrarily set to 1. Then, the full conditional of the last column of $\Omega$ is
	\begin{align*}
		p(\bm{\omega}_{(-p)p},{\omega}_{pp}\,|\, \Omega_{(-p)(-p)},Y,\Lambda,\tau) 
		\propto & ({\omega}_{pp}-\bm{\omega}_{(-p)p}'\Omega_{(-p)(-p)}^{-1}\bm{\omega}_{(-p)p})^{n/2} \\ & \times \text{exp}\{-\mathbf{s}_{(-p)p}'\bm{\omega}_{(-p)p}-{s}_{pp}{\omega}_{pp}/2-
		\bm{\omega}_{(-p)p}'(\Lambda^*\tau^2)^{-1}\bm{\omega}_{(-p)p}/2\},
	\end{align*}
	where $\Lambda^*$ is a diagonal matrix with $\bm{\lambda}_{(-p)p}$ in the diagonal.
	
	Next, a variable change is performed to obtain gamma and multivariate normal distributed variables, which can be efficiently sampled. Let $\bm{\beta}=\bm\omega_{(-p)p}$ and $\gamma={\omega}_{pp}-\bm{\omega}_{(-p)p}'\Omega_{(-p)(-p)}^{-1}\bm{\omega}_{(-p)p}$. The Jacobian of the transformation is a constant, and the full conditional of $\bm{\beta}$ and $\gamma$ is
	\begin{align}
		p(\bm{\beta},\gamma\,|\, \Omega_{(-p)(-p)},Y,\Lambda,\tau) &\propto \gamma^{n/2}\text{exp}[-\frac{1}{2}\{{s}_{pp}\gamma+
		\bm{\beta}'{s}_{pp}\Omega_{(-p)(-p)}^{-1}\bm{\beta}+\bm{\beta}'(\Lambda^*\tau^2)^{-1}\bm{\beta}+2\mathbf{s}_{(-p)p}'\bm{\beta}\}] \nonumber \\
		&\sim \text{Gamma}(n/2+1,{s}_{pp}/2)\text{Normal}(-C\mathbf{s}_{(-p)p},C), \label{eq:lik}
	\end{align}
	where $C=\{{s}_{pp}\Omega_{(-p)(-p)}^{-1}+(\Lambda^*\tau^2)^{-1}\}^{-1}$.
	
	Therefore the posterior distribution of the last row and column of $\Omega$ is obtained. All elements in the matrix $\Omega$ can be sampled by sampling one row and column at a time.
	
	Next, the local and global shrinkage parameters $\lambda_{ij}$ and $\tau$ need to be sampled. \citet{makalic2016samplerHS} made the following key observation: if
	$ x^2 \mid a \sim \mathrm{InvGamma} (1/2, 1/a)$ and $a\sim \mathrm{InvGamma} (1/2, 1)$, then marginally $x\sim C^{+} (0, 1)$, where the shape--scale parameterization is used for the inverse gamma distribution. The inverse gamma distribution is conjugate for the variance parameter in a linear regression model with normal errors and to itself, which ensures all required conditionals also follow inverse gamma distribution. 
	Thus, introduce latent $\nu_{ij}$ and write $\lambda_{ij}^2 \mid \nu_{ij} \sim \mathrm{InvGamma}(1/2, 1/\nu_{ij})$, and $\nu_{ij} \sim \mathrm{InvGamma}(1/2, 1)$. Since from Equation (\ref{eq:lik}), the full conditional posterior distribution of  $\bm\beta$ is normal, the full conditional posteriors of $\lambda_{ij}$ and $\nu_{ij}$ are easily obtained as $\lambda_{ij}^2 \mid \cdot \sim \mathrm{InvGamma}(1,1/\nu_{ij}+\omega_{ij}^2/2\tau^2)$ and $\nu_{ij}\mid \cdot \sim \mathrm{InvGamma}(1,1+1/\lambda_{ij}^2)$, respectively. Using a similar parameterization, the full conditional posteriors for $\tau^2$ and its auxiliary variable $\xi$ are also inverse gamma.

	Thus, combining the matrix partition and variable change for Bayesian graphical lasso proposed by \citet{wang2012bayesiangl} and the variable augmentation for the half-Cauchy prior proposed by \citet{makalic2016samplerHS}, the graphical horseshoe model has all conditionals in closed form and hence, admits a full Gibbs sampler. The sampler is summarized in Algorithm~\ref{alg:GHS}.
	
	\begin{algorithm}[!t]
	\caption{The Graphical Horseshoe Sampler}
	\label{alg:GHS}
	\begin{algorithmic}
		\Function{GHS}{$S,n,burnin,nmc$}
		\Comment{Where $S=Y'Y$, n=sample size}
		
		\State Set $p$ to be number of rows (or columns) in $S$
		\State Set initial values $\Omega = I_{p \times p},\, \Sigma = I_{p \times p},\, \Lambda = \mathbb{1},\, N = \mathbb{1},\, \tau=1,\, \xi=1$, where $\mathbb{1}$ is a matrix with all elements equal to 1, $\Lambda$ has entries of $\lambda_{ij}^2$, $N$ has entries of $\nu_{ij}$	
		
		\For{$iter = 1$ to $(burnin+nmc)$}
			\For{$i = 1$ to $p$}
				
				\State $\gamma \sim \text{Gamma}(\text{shape}=n/2+1,\, \text{rate}=2/{s}_{ii})$  \Comment{sample $\gamma$}
				\State $\Omega_{(-i)(-i)}^{-1} = \Sigma_{(-i)(-i)} - \bm{\sigma}_{(-i)i}\bm{\sigma}_{(-i)i}'/{\sigma}_{ii}$
				\State $C = ({s}_{ii}\Omega_{(-i)(-i)}^{-1}+\text{diag}(\bm{\lambda}_{(-i)i}\tau^2)^{-1})^{-1}$
				\State $\bm{\beta} \sim \text{Normal}(-C\mathbf{s}_{(-i)i},C)$  \Comment{sample $\bm{\beta}$}
				\State $\bm{\omega}_{(-i)i} = \bm{\beta},\, {\omega}_{ii} = \gamma + \bm{\beta}'\Omega_{(-i)(-i)}^{-1}\bm{\beta}$  \Comment{variable transformation}
				
				\State $\bm{\lambda}_{(-i)i} \sim \text{InvGamma}(\text{shape}=1,\, \text{scale}=1/ \bm{\nu}_{(-i)i}+\bm{\omega}_{(-i)i}^2/2\tau^2)$  \Comment{sample $\bm\lambda$, where $\bm{\lambda}_{(-i)i}$ is a vector of length $(p-1)$ with entries $\lambda_{ji}^2, j \ne i$}
				\State $\bm{\nu}_{(-i)i} \sim \text{InvGamma}(1,\, 1+1/ \bm{\lambda}_{(-i)i})$  \Comment{sample $\bm\nu$}
				
				\State Save updated $\Omega$
				\State $\Sigma_{(-i)(-i)}=\Omega_{(-i)(-i)}^{-1}+(\Omega_{(-i)(-i)}^{-1}\bm{\beta})(\Omega_{(-i)(-i)}^{-1}\bm{\beta})'/\gamma,\, \bm{\sigma}_{(-i)i}=-(\Omega_{(-i)(-i)}^{-1}\bm{\beta})/\gamma,\, {\sigma}_{ii}=1/\gamma$
				\State Save updated $\Sigma,\, \Lambda,\, N$				
			\EndFor
			\State $\tau^2 \sim \text{InvGamma}(({p\choose 2}+1)/2,\, 1/\xi+\sum_{i,j:i<j}{\omega_{ij}^2/2\lambda_{ij}^2})$  \Comment{sample $\tau$}
			\State $\xi \sim \text{InvGamma}(1,\, 1+1/\tau^2)$  \Comment{sample $\xi$}
		\EndFor
		
		\State Return MC samples $\Omega$
		
		\EndFunction
	\end{algorithmic}
	\end{algorithm}

	The constraint on $\Omega \in \mathcal{S}_p$ is maintained in every iteration as long as the starting value is positive definite, for the same reason that the positive definiteness is maintained in Bayesian graphical lasso \citep{wang2012bayesiangl}. Suppose that at iteration $t$, the current sample $\Omega^{(t)}$ is positive definite. Then all of its $p$ leading principal minors are positive. After updating the last column and row of $\Omega$, the new sample $\Omega^{(t+1)}$ has the same leading principal minors as $\Omega^{(t)}$ except for the last one which is of order $p$. The last leading principal minor is $\text{det}(\Omega^{(t+1)})=\gamma \text{det}(\Omega_{(-p)(-p)}^{(t)})$, and is positive since both $\gamma$ and $\text{det}(\Omega_{(-p)(-p)}^{(t)})$ are positive. Consequently, $\Omega^{(t+1)}$ after updating is positive definite.
	
  The required full conditionals in the proposed Gibbs sampler are either multivariate normal, gamma or inverse gamma, for which efficient sampling methods exist. Full conditional posteriors of the local shrinkage parameters $\lambda_{ij:i<j}$ are mutually independent, and so are $\nu_{ij:i<j}$. This facilitates batch updating and a large number of features does not cause problems in sampling of $\bm\lambda$ and $\bm\nu$. The most computationally expensive step is the sampling of $\bm{\beta}$, where the $(p-1)\times (p-1)$ matrix $\Omega_{(-p)(-p)}$ and $\{{s}_{pp}\Omega_{(-p)(-p)}^{-1}+(\Lambda^*\tau^2)^{-1}\}$ need to be inverted, which has computational complexity $O(p^3)$. In Algorithm~\ref{alg:GHS}, $\Omega_{(-p)(-p)}$ is inverted by block form of the sampled covariance matrix, so only $\{{s}_{pp}\Omega_{(-p)(-p)}^{-1}+(\Lambda^*\tau^2)^{-1}\}$ needs to be inverted. MATLAB code for Algorithm~\ref{alg:GHS}, along with a simulation example, are freely available at \textit{http://github.com/liyf1988/GHS}.

\section{Kullback--Leibler Risk Bounds}
\label{sec:KL}
	
	In this section, we discuss the Kullback--Leibler divergence between the true sampling density and the Bayes estimator of the density function under various priors, including the graphical horseshoe prior. The Ces\`aro-average risk of the posterior distribution diverges for all methods when $p^2/n \to \infty$, but the upper bound of the average risk under the graphical horseshoe prior diverges slower than some other methods, as discussed below.
	
	Suppose that there is a true sampling model. Let $\Omega_0$ denote the true value of the precision matrix, $p_{\Omega}=p(y\,|\, \Omega)$ denote a sampling density with parameter $\Omega$, and $\nu(A)$ denote the measure of some set $A$. Let $D(p_0||p_1)=\mathrm{E}_{p_2}{\text{log}\,(p_2/p_1)}$ denote the Kullback--Leibler divergence from $p_1$ to $p_2$. Then \cite{barron1988exponential} proved the following lemma on the Ces\`aro-average risk of the Bayes posterior mean estimator of the density function.
	
	\begin{lemma}
	\label{lm:information}
	\citep{barron1988exponential} Let $A_{\epsilon}=\{\Omega: D(p_{\Omega_0} || p_{\Omega}) \leq \epsilon\} \subset \mathbb{R}^{p\times p}$ denote the Kullback--Leibler information neighborhood of size $\epsilon$, centered at $\Omega_0$. Let $\nu(d\Omega)$ be the prior measure of $\Omega$ and $\nu_n(d\Omega) \propto \prod_{i=1}^{n} p_\Omega(y_i) \nu(\mathrm{d}\Omega)$ be the posterior measure after observing i.i.d. $y_1,...,y_n$ from the sampling density $p_\Omega$. Let $\hat{p}_n=\int p_{\Omega} \nu_n(d\Omega)$ be the posterior mean estimator of the density function.
	Under the assumption that the prior measure $\nu(A_{\epsilon})>0$ for all $\epsilon>0$, the Ces\`aro-average risk $R_n$ of the estimator $\hat{p}_n$ admits the following upper bound for all $\epsilon>0$:
	\begin{equation*}
		R_n = \frac{1}{n}\sum_{j=1}^n \mathrm{E} D(p_{\Omega_0} || \hat{p}_j) \leq \epsilon-\frac{1}{n}\textnormal{log} \, \nu(A_{\epsilon}),
	\end{equation*}
	where the expectation is with respect to the posterior predictive distribution given $y_1, ..., y_n$.
	\end{lemma}

	Taking $\epsilon=1/n$, the upper bound of $R_n$ is a function of two things: the sample size $n$, and the prior measure of the Kullback--Leibler information neighborhood $A_{\epsilon}$ of true $\Omega_0$. Since the horseshoe prior has higher mass near the true parameter than any prior that is bounded above when the true parameter is zero, the graphical horseshoe estimator has a smaller upper bound on $R_n$ when the true precision matrix is sparse. The result is summarized in the following theorem.
	
	\begin{theorem}
	\label{thm:bound}
	Suppose the true sampling model is $y \sim \textnormal{Normal}(\mathbf{0}, \Omega_0)$. Let $\sigma_{ij0}$ denote the $ij$th element of the true covariance matrix $\Sigma_0$, and $\omega_{ij0}$ denote the $ij$th element of the true precision matrix $\Omega_0$. Suppose that $\sum_{i,j}\sigma_{ij0}=Mp$ where M is a constant. That is, the summation of all elements in $\Sigma_0$ grows linearly with the number of features $p$. Suppose that an Euclidean cube in the neighborhood of $\Omega_0$ with $(\omega_{ij0}-2/Mn^{1/2}p, \omega_{ij0}+2/Mn^{1/2}p)$ on each dimension lies in the cone of positive definite matrices $\mathcal{S}_p$. Then $ R_n \leq \frac{1}{n}-\frac{1}{n}\textnormal{log} \, \nu(A_{1/n})$ for all $n$, and:
	
	(1) For $\hat{p}_n$ under the graphical horseshoe prior, {$ \textnormal{log} \, \nu(A_{1/n}) > p_0 \textnormal{log} \, \{ \frac{C_1}{Mn^{1/2}p} \textnormal{log} \, (2Mn^{1/2}p) \} + p_1 \textnormal{log} \, \frac{C_2}{n^{1/2}p} $}, where $p_0$ is the number of zero elements in $\Omega_0$, $p_1$ is the number of nonzero elements in $\Omega_0$, and $C_1$ and $C_2$ are constants.
	
	(2) Suppose $p(\omega_{ij})$ is any other prior density that is continuous, bounded above, and strictly positive on a neighborhood of the true value $\omega_{ij0}$. Then $\textnormal{log} \, \nu(A_{1/n}) > p^2 \textnormal{log} \, \frac{K_1}{n^{1/2}p} $, where $K_1$ is a constant.
	
	\end{theorem}
	
	Proof of Theorem~\ref{thm:bound} can be found in Supplementary Material. The neighborhood $A_{1/n}$ is bounded by two Euclidean cubes on $p\times p$ dimensions where the edges of these cubes have length proportional to $n^{1/2}p$ on each dimension. On these cubes, the measure of $p(\Omega)$ is obtained by the product of the measures of $p(\omega_{ij})$ on each of the $p^2$ dimensions of $\Omega$. Any Bayesian estimator with a prior density bounded above near the origin gives a prior measure of order $1/(n^{1/2}p)$ on each dimension. The graphical horseshoe estimator gives a prior measure of order $\text{log} \, (n^{1/2}p)/(n^{1/2}p)$ on each dimension with $\omega_{ij0}=0$, and a measure of order $1/(n^{1/2}p)$ on each dimension with nonzero $\omega_{ij0}$.
	
	Some common Bayesian estimators, including the double exponential prior in Bayesian lasso, induce a prior density bounded above near the origin \citep{carvalho2010horseshoe}. Although the SCAD estimate can not be expressed as a maximum a posteriori estimate, the prior density corresponding to the SCAD penalty would be bounded by Theorem 1 in \citet{polson2012levy}. Therefore, Bayesian graphical lasso has an upper bound corresponding to Part (2) of Theorem~\ref{thm:bound}. Similarly, the posterior distribution of a Bayesian version of graphical SCAD would also have an upper bound corresponding to Part (2) of Theorem~\ref{thm:bound}, if such a Bayesian version existed. These methods put a prior measure of order $1/(n^{1/2}p)$ near the true parameter on each dimension, regardless of whether or not the true parameter is zero. Unlike the horseshoe prior, these methods do not put dense prior mass near the origin, and do not utilize the fact (or expectation) that most of the true parameters are zero.
	
	Theorem 1 of \cite{rissanen1986stochastic} gives an asymptotic lower bound on the Kullback--Leibler divergence $D(p_{\Omega_0} || \hat{p}_n)$, which is $(1/2-\epsilon)\, k \, \textnormal{log}\, n$ for all $\epsilon>0$, where $k$ is the dimension of the parameter vector. This lower bound implies that in our problem, all methods have divergent average risk $R_n$ when $n \to \infty$ and $p^2/n \to \infty$.
	Though all methods fail when dimension is large, Theorem~\ref{thm:bound} gives a non-asymptotic upper bound on $R_n$ for any sample size $n$. One element where the true parameter is zero contributes $(\text{log} \, n^{1/2}p - \text{log} \, C)/n$ to the upper bound of $R_n$ under a bounded prior near the origin, and $(\text{log}\,Mn^{1/2}p - \text{log}\,C - \text{log}\,\text{log}\,2Mn^{1/2}p)/n$ to the upper bound of $R_n$ under the graphical horseshoe prior. For each element where the true parameter is zero, the graphical horseshoe average risk has an extra $-O\{(\text{log}\,\text{log}\,n^{1/2}p)/n\}$ term. Consequently, when most off-diagonal elements in the true precision matrix are zero, the graphical horseshoe estimate provides a non-trivial improvement on $R_n$. In Section~\ref{sec:sim}, we will compare the Kullback--Leibler divergence of graphical horseshoe estimate to graphical lasso and graphical SCAD estimates in simulations. We will show that the graphical horseshoe estimate has smaller Kullback--Leibler divergence, especially when the precision matrix is extremely sparse. However, we will discuss the bias of the graphical horseshoe estimate first.
	
\section{Bias of the Graphical Horseshoe Estimator}
\label{sec:bias}


Suppose that all diagonal elements in the precision matrix are known. Then, by the partial regression representation of the parameters \citep{pourahmadi2011covariance}, given the rest of the features, an observation of the $i$th feature follows a normal distribution $y_i\,|\, \mathbf{y}_{(-i)} \sim \text{Normal}(-\omega_{ii0}^{-1}\bm{\omega}_{i(-i)0}\mathbf{y}_{(-i)},\omega_{ii0}^{-1})$, where $y_i$ is an observation of the $i$th feature, $\mathbf{y}_{(-i)}$ is an observation of all features other than $i$, $\omega_{ii0}$ is the diagonal element in the true precision matrix corresponding to feature $i$, and $\bm{\omega}_{i(-i)0}$ is the off-diagonal elements in the true precision matrix on the $i$th row. Without loss of generality, the following discussion takes $i=p$. Given observations of features $1$ to $p-1$, $Y_{(-p)}$, the least squares estimate of the $p$th column in the precision matrix is an unbiased estimate with a normal distribution
	\begin{equation*}
		\hat{\bm{\omega}}_{p(-p)}\, | \, Y_{(-p)} \sim \text{Normal}(\bm{\omega}_{p(-p)0}, w_{pp0}(Y_{(-p)}'Y_{(-p)})^{-1}).
	\end{equation*}
	Marginally, the least squares estimate of each element $\hat{\omega}_{pj}$ in $\hat{\bm{\omega}}_{p(-p)}$ has a univariate normal distribution	
	\begin{equation*}
		\hat{\omega}_{pj}\, | \, Y_{(-p)} \sim \text{Normal}(\omega_{pj0}, w_{pp0}(Y_{(-p)}'Y_{(-p)})_{jj}^{-1}).
	\end{equation*}
We use this property of the least squares estimate to state our main result on the element-wise bias of the graphical horseshoe estimate.
	
	\begin{theorem}
	\label{thm:bias}
	Scale both $\omega_{pj}$ and its least squares estimate $\hat{\omega}_{pj}$ by $\{\omega_{pp0}(Y_{(-p)}'Y_{(-p)})_{jj}^{-1}\}^{-1/2}$, and denote the scaled parameter and its least squares estimate by $\omega_{pj}'$ and $\hat{\omega}_{pj}'$, respectively.	
	
	(1) The posterior mean estimate of $\omega_{pj}'$ under the graphical horseshoe prior is $\mathrm{E}(\omega_{pj}'\, | \, Y, \tau)=(1-\mathrm{E}(Z_{pj}))\hat{\omega}_{pj}'$, where the random variable $Z_{pj}$ follows a Compound Confluent Hypergeometric (CCH) distribution with parameters $(1,1/2,1,\hat{\omega}_{pj}'^2/2,1,\omega_{pp0}(Y_{(-p)}'Y_{(-p)})_{jj}^{-1}\tau^{-2})$ and has support between $0$ and $1$. Let $\theta_{pj}=\omega_{pp0}(Y_{(-p)}'Y_{(-p)})_{jj}^{-1}\tau^{-2}$, then $\mathrm{E}(Z_{pj}) < 4(C_1+C_2)\theta_{pj}(1+\hat{\omega}_{pj}'^2/2)/\hat{\omega}_{pj}'^4$ when $\hat{\omega}_{pj}'^2/2>1$, where $C_1=1-2e \approx 0.26$ and $C_2=\Gamma(1/2)\Gamma(2)/\Gamma(2.5)=0.75$. Consequently, $\mathrm{E}(Z_{pj})=O(1/\hat{\omega}_{pj}'^2)$ when $\hat{\omega}_{pj}' \to \infty$.
	
	(2) The posterior mean estimate of $\omega_{pj}'$ under the double-exponential prior is $\mathrm{E}(\omega_{pj}'\, | \, Y)_{lasso} = \hat{\omega}_{pj}'+\frac{\mathrm{d}}{\mathrm{d}\hat{\omega}_{pj}'}\mathrm{log} \, m_{lasso}(\hat{\omega}_{pj}')$, where $m_{lasso}(\hat{\omega}_{pj}')$ is the marginal distribution of $\hat{\omega}_{pj}'$ under the double-exponential prior. Moreover, $lim_{|\hat{\omega}_{pj}'|\to\infty} \frac{\mathrm{d}}{\mathrm{d}\hat{\omega}_{pj}'}\mathrm{log} \, m_{lasso}(\hat{\omega}_{pj}')=\pm a$, where $a=2^{1/2}/nv$ and $v$ is the variance of the double-exponential prior.
	
	(3) The squared scaled least squares estimate follows a noncentral Chi-squared distribution with one degree of freedom, i.e. $\hat{\omega}_{pj}'^2\, | \, Y_{(-p)} \sim \text{Noncentral} \,  \chi^2(1, \omega_{pj}'^2)$, and by the scaling, $\omega_{pj}'^2=\omega_{pj0}^2\omega_{pp0}^{-1}\{(Y_{(-p)}'Y_{(-p)})_{jj}^{-1}\}^{-1}$.  When $n>p-1$, $\{(Y_{(-p)}'Y_{(-p)})_{jj}^{-1}\}^{-1} \sim \text{Gamma}((n-p+2)/2, 2(\omega_{jj0}-\omega_{pj0}^2/\omega_{pp0})^{-1})$.
	\end{theorem}
	
	Proof of Theorem~\ref{thm:bias} is in our Supplementary Material. A very brief introduction to the CCH distribution and the upper bound of $\mathrm{E}(Z)$, where $Z \sim CCH(1,1/2,1,s,1,\theta)$, can be found in \citet{bhadra2016prediction}. Part (1) of Theorem~\ref{thm:bias} states that given the data, the element-wise graphical horseshoe estimate is close to (in fact $O(1/\hat{\omega}_{pj}')$ away from) the unbiased least squares estimate when $\hat{\omega}_{pj}'^2$ is large, for any fixed global shrinkage parameter $\theta_{pj}$. This property of the posterior mean is a consequence of the half-Cauchy distribution in the horseshoe prior. One may notice that the parameter $\theta_{pj}$ in the CCH distribution depends on the data. However, the global shrinkage parameter $\tau$ can be estimated to control $\theta_{pj}$ and $\mathrm{E}(Z_{pj})$, so that the graphical horseshoe estimate has the desired shrinkage.
	
	On the other hand, Part (2) of Theorem~\ref{thm:bias} asserts that the posterior mean estimate of Bayesian graphical lasso does not converge to the unbiased least squares estimate for any finite $n$, even when $\hat{\omega}_{pj}'$ is large. In addition, the term $a$ varies inversely with the global shrinkage parameter in the double-exponential prior and tends to be large in sparse cases \citep{carvalho2009handling}. Therefore, in sparse cases, the posterior mean estimate of Bayesian graphical lasso tends to be further away from the unbiased least squares estimate.
	
	Part (3) of Theorem~\ref{thm:bias} implies the condition that $\hat{\omega}_{pj}'^2$ is large is met with high probability when sample size is large. The parameter $\hat{\omega}_{pj}'^2$ has a noncentral $\chi^2$ distribution with noncentrality parameter $\omega_{pj}'^2$ and $1$ degree of freedom. The noncentrality parameter $\omega_{pj}'^2$ equals to a constant $\omega_{pj0}^2\omega_{pp0}^{-1}$ times a gamma distributed variable. This gamma distributed variable $\{(Y_{(-p)}'Y_{(-p)})_{jj}^{-1}\}^{-1}$ has mean proportional to $n-p+2$ and mode proportional to $n-p$. Therefore, when $\omega_{pj0} \neq 0$ and $n \gg p$, both $\omega_{pj}'^2$ and $\hat{\omega}_{pj}'^2$ are large with high probability, and the graphical horseshoe estimate of $\omega_{pj}'$ is $O(1/\hat{\omega}_{pj}')$ away from the unbiased least squares estimate. 
	
	To summarize the main implications of Theorem ~\ref{thm:bias}, when $n \gg p$ and the true parameter is nonzero, the graphical horseshoe estimate is close to an unbiased estimator with high probability, while the posterior mean estimate of Bayesian graphical lasso is not. When the sample size is sufficiently large, the bias of graphical horseshoe estimate is low for nonzero elements even though the method shrinks zero elements heavily. The theorem depends on the least squares estimate, which does not exist when $n<p$. However, graphical horseshoe is a shrinkage method that gives a stable estimate even when $n<p$. The bias of graphical horseshoe estimate is affected by the constant $\omega_{pj0}^2 \omega_{pp0}^{-1}$, which implies that bias would be small when $\omega_{pj0}^2 \omega_{pp0}^{-1}$ is large even in a $n<p$ case. In Section~\ref{sec:sim}, we numerically demonstrate the error of graphical horseshoe estimates under various situations, with both $n>p$ and $n<p$.
	
\section{Simulation Study}
\label{sec:sim}
	
	In this section, simulations are performed to compare the graphical horseshoe, graphical lasso, graphical SCAD, and Bayesian graphical lasso estimators. In the first example, we consider $p=100$ features and $n=50$ observations. The precision matrix $\Omega_0$ is taken to be sparse with diagonal elements set to one and one of the following three patterns for off-diagonal elements \citep{friedman2010applications}:
	
	\textit{Random.} Each off-diagonal element is randomly set to $\omega_{ij}<0$ (corresponding to positive partial correlations) with probability 0.01, where the magnitude of nonzero off-diagonal elements is uniformly selected between $-1$ and $-0.2$. For these simulations, we consider 35 nonzero elements in $\Omega_0$ with values ranging between $-0.8397$ and $-0.2044$.
	
	\textit{Hubs.} The rows/columns are partitioned into disjoint groups $\{G_k\}_1^K$. Each group has a row $k$ where off-diagonal elements are taken to be $\omega_{ik}=0.25$ (corresponding to negative partial correlations) for $i \in G_k$ and $\omega_{ij}=0$ otherwise. We consider $10$ groups and $10$ members within each group, giving $90$ nonzero off-diagonal elements in $\Omega_0$.
	
	\textit{Cliques.} The rows/columns are partitioned into disjoint groups and $\omega_{ij:i,j \in G_k,\, i\neq j}$ are set to $-0.45$ corresponding to a positive partial correlation case and to $0.75$ corresponding to a negative partial correlation case. We again consider $10$ groups but only three members within each group, resulting in $30$ nonzero off-diagonal elements in $\Omega_0$.

	In our second and third examples, we consider $p=100, n=120$ and $p=200, n=120$, using the sparsity structures above. The $p=100, n=120$ case uses the same precision matrix as the $p=100, n=50$ case. For the $p=200, n=120$ case, all settings for the precision matrix are kept the same except in the random structure where each off-diagonal $\omega_{ij}$ has a probability of 0.002 of being nonzero.  The results for the three examples are summarized in Tables~\ref{tab:1}, \ref{tab:2} and \ref{tab:3}, respectively. 
	
	For each choice of $\Omega_0$, 50 data sets are generated and $\Omega$ is estimated using the graphical SCAD, graphical horseshoe, frequentist graphical lasso with and without penalization on diagonal elements, and the Bayesian graphical lasso. Our graphical horseshoe estimator is implemented in \citet{MATLAB:2018}. We use the posterior mean as our estimate. MATLAB code by \citet{wang2012bayesiangl} is used for the graphical SCAD and Bayesian graphical lasso. The frequentist graphical lasso is implemented using the package ``glasso'' \citep{R:glasso} in R \citep{R}. Tuning parameters in the graphical lasso and graphical SCAD are selected by five-fold cross validation using log likelihood. In the case where $p=100$ and $n=120$, an estimate of $\Omega$ based on the unpenalized likelihood function is feasible, and we also include a refitted graphical lasso in this comparison. For the refitted graphical lasso, the graphical lasso is first applied for variable selection, then the selected parameters in $\Omega$ are refitted using the graphical lasso algorithm, with the tuning parameter fixed at zero (i.e. no penalization). For the refitted graphical lasso, log likelihood of the final unpenalized estimate is used to calculate the cross validation score, used in selecting the tuning parameter in the variable selection step.
	
	Stein's loss of the estimated precision matrix $\Omega$ (which equals to 2 times the Kullback--Leibler divergence of $\Omega$ from $\Omega_0$), Frobenius norm of $\Omega-\Omega_0$, true positive rate (TPR), and false positive rate (FPR) are calculated. Since both graphical SCAD and graphical lasso provide variable selection in their estimates (i.e., some of the elements are estimated to be zero), their variable selection results are calculated using the number of nonzero estimates. Graphical horseshoe and the Bayesian graphical lasso, however, are shrinkage methods and do not estimate elements to be exactly equal to zero. For these two methods, we use the symmetric central $50\%$ posterior credible intervals for variable selection. That is, if the $50\%$ posterior credible interval of an off-diagonal element of $\Omega$ does not contain zero, that element is considered a discovery, and vice versa. For each statistic, we report the mean and standard deviation computed over 50 data sets. We also report the average CPU time in minutes for each method. We provide additional simulation results, a larger dimensional setting with $p=400$ and $n=120$, and MCMC convergence diagnostics in the Supplementary Material. The simulations were performed on a server with 1TB of RAM, and 20 total CPU cores from a pair of Intel Xeon E5-2660 v3 CPUs at 2.60GHz, with 10 cores each.

	\subsection{Estimation}
	\label{sec:estimation}
	
 From Tables~\ref{tab:1}, \ref{tab:2} and \ref{tab:3}, the graphical horseshoe estimate has the smallest Stein's loss and the smallest Frobenius norm (F norm) among the regularization methods considered, in eleven and ten out of twelve cases, respectively. When $p=100$ and $n=120$, an estimation of $\Omega$ based on the unpenalized likelihood is feasible, since $n>p$. In this case, the refitted graphical lasso, based on variable selection by graphical lasso and unpenalized estimation of the selected variables, performs well (Table~\ref{tab:2}). However, the graphical horseshoe performs comparably to the refitted graphical lasso, except for the hubs structured precision matrix. The graphical horseshoe is expected to perform well when the precision matrix is sparse and the absolute values of scaled nonzero elements are large. In our simulations, the hubs structure is the least sparse with small nonzero elements, and the cliques structured matrix with negative partial correlations is the sparsest with larger nonzero elements. Simulation results confirm that the advantage of graphical horseshoe is indeed larger in the cliques structure with negative partial correlations, and smaller in the hubs structure, if there is an advantage at all.

	In the simulations, the graphical SCAD and frequentist graphical lasso with penalized diagonal terms are comparable in terms of Stein's loss and Frobenius norm. The frequentist graphical lasso with unpenalized diagonal terms performs somewhat worse. The Bayesian graphical lasso is by far the worst in estimation, especially in terms of Stein's loss, in accordance with the results in Section~\ref{sec:KL}.
	
\begin{figure}[t]
	\centering
	\begin{subfigure}[b]{0.325\textwidth}
		\centering
		\adjincludegraphics[width=\textwidth,trim={{0.02\width} {0\height} {0.05\width} {0\height}},clip]{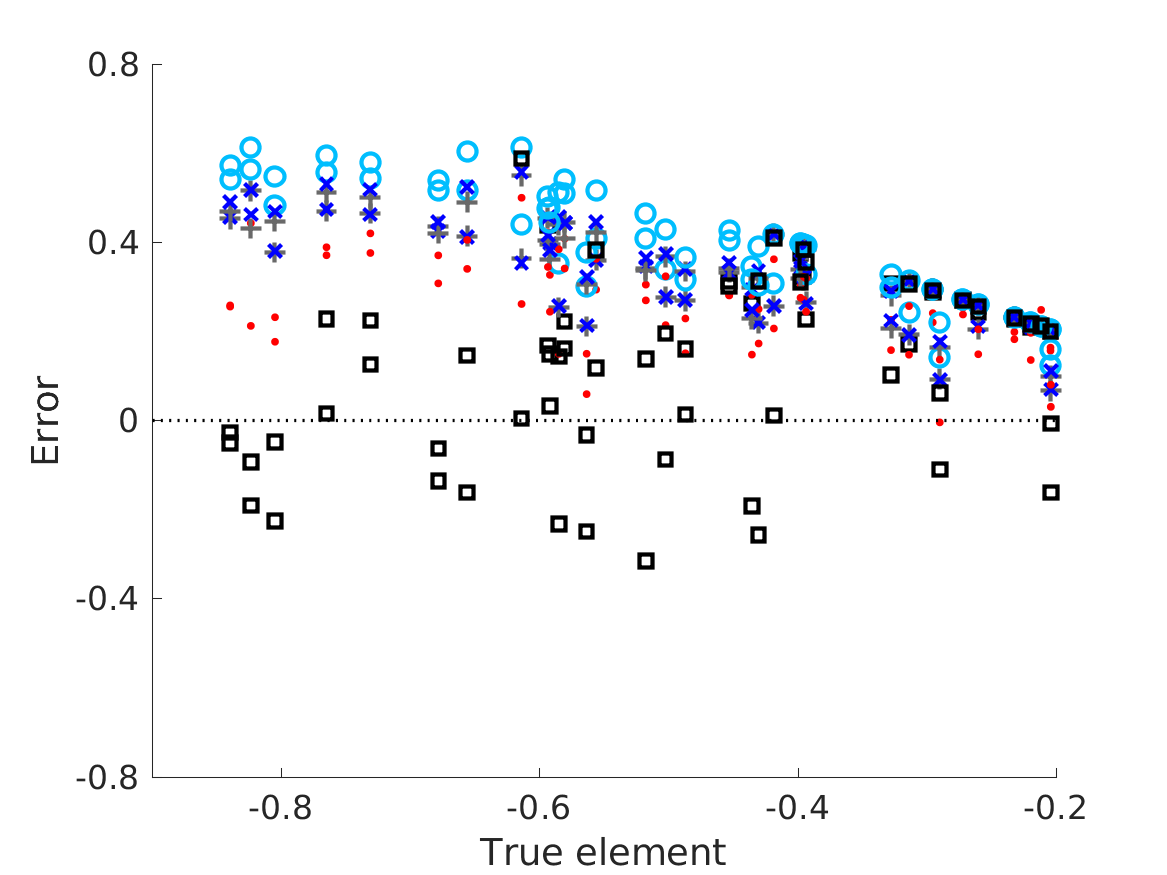}
		\caption[Bias]%
		{{\footnotesize $p=100, n=50$}}
		\label{fig:Bias_p100random}
	\end{subfigure}
	\
	\begin{subfigure}[b]{0.325\textwidth}
		\centering
		\adjincludegraphics[width=\textwidth,trim={{0.02\width} {0\height} {0.05\width} {0\height}},clip]{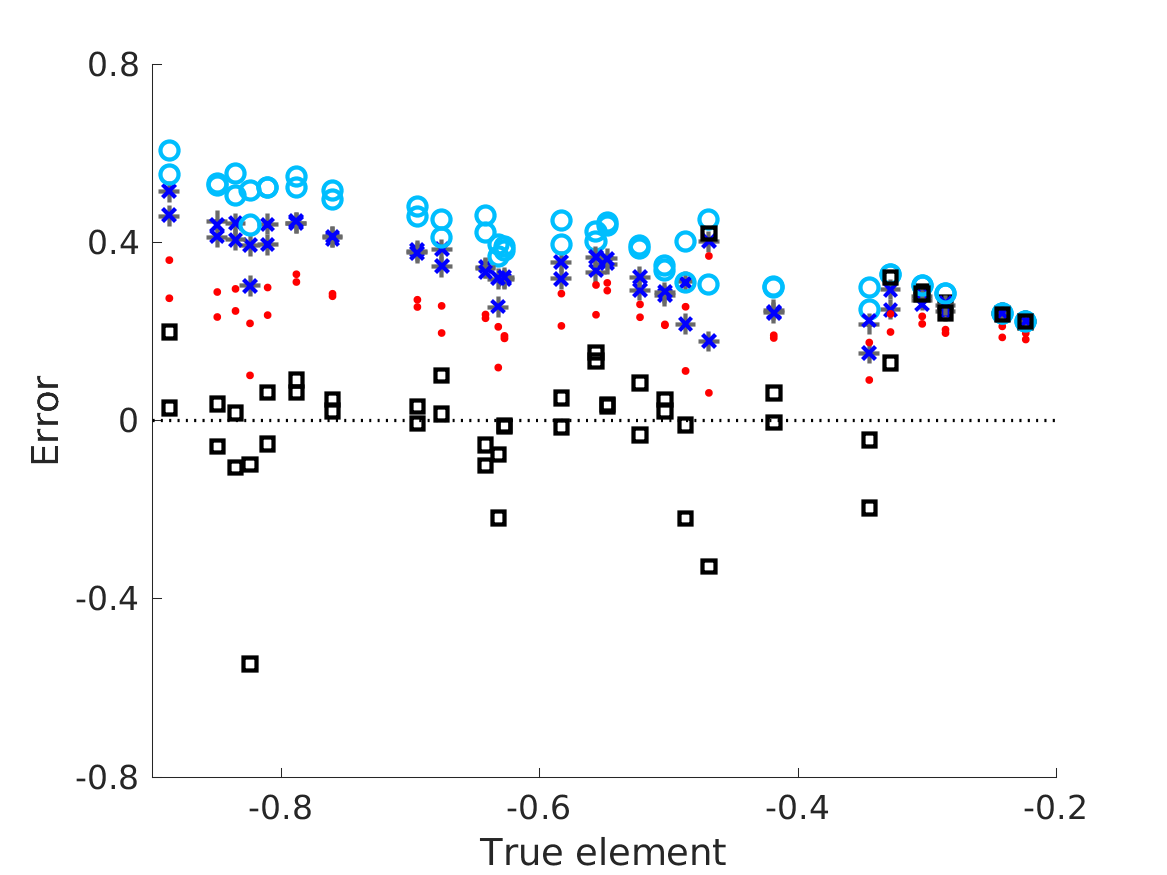}
		\caption[]%
		{{\footnotesize $p=200, n=120$}}
		\label{fig:Bias_p200randomn120}
	\end{subfigure}
	\
	\begin{subfigure}[b]{0.325\textwidth}
		\centering
		\adjincludegraphics[width=\textwidth,trim={{0.02\width} {0\height} {0.05\width} {0\height}},clip]{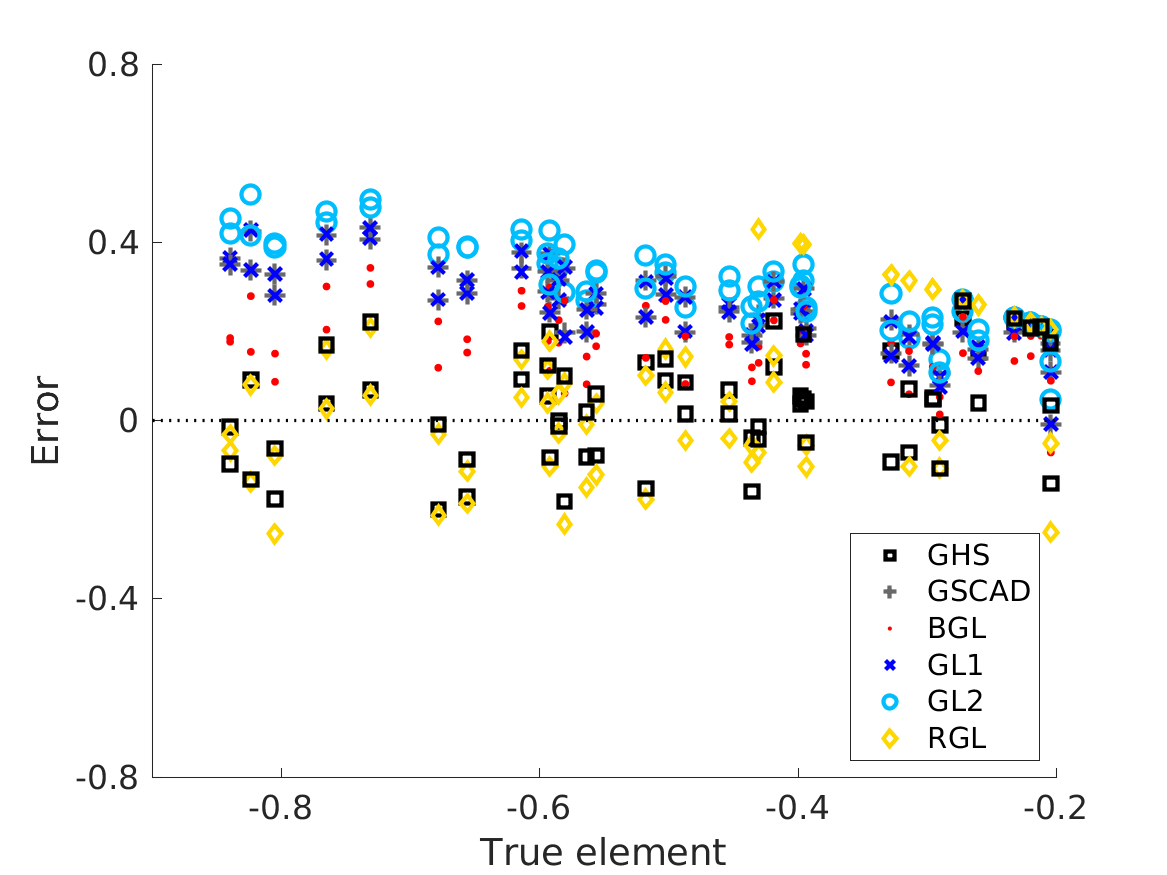}
		\caption[]%
		{{\footnotesize $p=100, n=120$}}
		\label{fig:Bias_p100randomn120}
	\end{subfigure}
	\caption{Errors of nonzero elements of estimated precision matrix by frequentist graphical lasso with penalized diagonal elements (GL1), frequentist graphical lasso with unpenalized diagonal elements (GL2), refitted graphical lasso (RGL), graphical SCAD (GSCAD), Bayesian graphical lasso (BGL), and graphical horseshoe (GHS). Random structure of precision matrix. Estimates using two representative data sets in simulations.}
	\label{fig:bias}
\end{figure}

	Figure~\ref{fig:bias} shows the estimation errors in nonzero off-diagonal elements for the random structured precision matrix. As a plot showing estimation errors using all 50 data sets will be hard to read, errors in only two representative data sets in simulations are shown in each plot. Scatterplots indicate that the errors in the graphical horseshoe estimates are randomly scattered around zero while the graphical lasso, graphical SCAD and Bayesian graphical lasso always shrink the estimates toward zero. When $p=100$ and $n=120$, graphical horseshoe estimates have errors comparable to the unpenalized refitted graphical lasso errors. Graphical horseshoe estimates also have smaller errors than the other estimates, especially when absolute values of true elements are large and when $n-p$ is large. These results agree with the theory and discussion in Section~\ref{sec:bias}.

\begin{table}[!t]
	\centering
	\caption{Mean (sd) Stein's loss, Frobenius norm, true positive rates and false positive rates of precision matrix estimates over 50 data sets generated by multivariate normal distributions with precision matrix $\Omega_0$, where $p=100$ and $n=50$. The precision matrix is estimated by frequentist graphical lasso with penalized diagonal elements (GL1), frequentist graphical lasso with unpenalized diagonal elements (GL2), graphical SCAD (GSCAD), Bayesian graphical lasso (BGL), and graphical horseshoe (GHS). The best performer in each row is shown in bold. Average CPU time is in minutes.}
	
	\label{tab:1}
	\begin{footnotesize}
		\noindent\makebox[\textwidth]{%
			\begin{tabular}{l|rrrrr|rrrrr|}
				\toprule
				& \multicolumn{5}{|c|}{Random} & \multicolumn{5}{|c|}{Hubs} \\
				nonzero pairs & \multicolumn{5}{|c|}{35/4950} & \multicolumn{5}{|c|}{90/4950} \\
				nonzero elements & \multicolumn{5}{|c|}{$\sim -\mathrm{Unif}(0.2,1)$} & \multicolumn{5}{|c|}{0.25} \\
				$p=100, n=50$ & GL1 & GL2 & GSCAD & BGL & GHS & GL1 & GL2 & GSCAD & BGL & GHS \\
				\toprule   	
				Stein's loss & 10.20 & 13.42 & 10.05 & 80.92 & \textbf{6.44} 
				& 10.12 & 12.78 & \textbf{10.01} & 77.85 & 12.56 \\
				& (0.53) & (1.06) & (0.55) & (1.63) & (0.85)
				& (0.53) & (0.96) & (0.50) & (1.66) & (1.04)\\
				
				F norm & 4.33 & 5.30 & 4.31 & 5.58 & \textbf{3.31}
				& 3.95 & 4.63 & \textbf{3.94} & 5.97 & 3.96 \\
				& (0.18) & (0.20) & (0.16) & (0.26) & (0.29)
				& (0.13) & (0.18) & (0.14) & (0.30) & (0.27) \\
				
				TPR & .8246 & .7097 & \textbf{.9977} & .8709 & .5903 
				& .8649 & .7333 & \textbf{.9987} & .8513 & .2687 \\		
				& (.0520) & (.0620) & (.0078) & (.0470) & (.0537)
				& (.0443) & (.0751) & (.0053) & (.0378) & (.0764) \\
				
				FPR & .0947 & .0374 & .9955 & .1055 & \textbf{.0004}
				& .0919 & .0281 & .9976 & .1189 & \textbf{.0013} \\
				& (.0141) & (.0070) & (.0102) & (.0059) & (.0003)
				& (.0130) & (.0086) & (.0069) & (.0058) & (.0005)\\
				
				Avg CPU time & 0.30 & 0.35 & 6.24 & 40.94 & 38.32
				& 0.14 & 0.16 & 4.01 & 35.44 & 41.58 \\
				
				\toprule
				& \multicolumn{5}{|c|}{Cliques positive} & \multicolumn{5}{|c|}{Cliques negative} \\
				nonzero pairs & \multicolumn{5}{|c|}{30/4950} & \multicolumn{5}{|c|}{30/4950} \\
				nonzero elements & \multicolumn{5}{|c|}{-0.45} & \multicolumn{5}{|c|}{0.75} \\
				$p=100, n=50$ & GL1 & GL2 & GSCAD & BGL & GHS & GL1 & GL2 & GSCAD & BGL & GHS \\
				\toprule
				Stein's loss & 9.16 & 14.16 & 8.99 & 81.58 & \textbf{5.87}
				& 11.00 & 14.37 & 10.90 & 81.27 & \textbf{6.28} \\
				& (0.55) & (1.06) & (0.52) & (2.51) & (0.93)
				& (0.43) & (1.02) & (0.43) & (1.98) & (1.09) \\
				
				F norm 	& 3.75 & 5.01 & \textbf{3.71} & 5.44 & 3.81
				& 6.00 & 6.86 & 5.99 & 6.51 & \textbf{3.64} \\
				& (0.16) & (0.16) & (0.17) & (0.33) & (0.41)
				& (0.14) & (0.16) & (0.14) & (0.20) & (0.36) \\
				
				TPR & \textbf{1} & \textbf{1} & \textbf{1} & \textbf{1} & .7487 
				& .9993 & .9880 & \textbf{1} & .9993 & .9733 \\
				& (0) & (0) & (0) & (0) & (.0427)
				& (.0047) & (.0221) & (0) & (.0047) & (.0421) \\
				
				FPR & .0900 & .0255 & .9901 & .1014 & \textbf{.0003} 
				& .0922 & .0279 & .9752 & .1161 & \textbf{.0010} \\
				& (.0098) & (.0056) & (.0177) & (.0052) & (.0003)
				& (.0135) & (.0084) & (.0219) & (.0051)  & (.0005) \\
				
				Avg CPU time & 0.24 & 0.28 & 4.52 & 34.45 & 41.65
				& 0.18 & 0.20 & 6.91 & 33.88 & 41.05 \\	   
				\bottomrule
		\end{tabular}}
	\end{footnotesize}
\end{table}

\begin{table}[!t]
	\centering
	\caption{Mean (sd) Stein's loss, Frobenius norm, true positive rates and false positive rates of precision matrix estimates over 50 data sets generated by multivariate normal distributions with precision matrix $\Omega_0$, where $p=100$ and $n=120$. The precision matrix is estimated by frequentist graphical lasso with penalized diagonal elements (GL1), frequentist graphical lasso with unpenalized diagonal elements (GL2), refitted graphical lasso (RGL), graphical SCAD (GSCAD),  Bayesian graphical lasso (BGL), and graphical horseshoe (GHS). The best performer in each row is shown in bold. Average CPU time is in minutes.}
	
	\label{tab:2}
	\begin{footnotesize}
		\noindent\makebox[\textwidth]{%
			\addtolength{\tabcolsep}{-3pt}
			\begin{tabular}{l|rrrrrr|rrrrrr|}
				\toprule
				& \multicolumn{6}{|c|}{Random} & \multicolumn{6}{|c|}{Hubs} \\
				nonzero pairs & \multicolumn{6}{|c|}{35/4950} & \multicolumn{6}{|c|}{90/4950} \\
				nonzero elements & \multicolumn{6}{|c|}{$\sim -\mathrm{Unif}(0.2,1)$} & \multicolumn{6}{|c|}{0.25} \\
				$p=100, n=120$ & GL1 & GL2 & RGL & GSCAD & BGL & GHS & GL1 & GL2 & RGL & GSCAD & BGL & GHS \\
				\toprule   	
				Stein's loss & 5.32 & 6.90 & 3.84 & 5.29 & 43.08 & \textbf{2.15} 
				& 5.34 & 6.53 & \textbf{3.92} & 5.29 & 43.07 & 5.12 \\
				& (0.27) & (0.51) & (0.46) & (0.26) & (0.82) & (0.27)
				& (0.28) & (0.47) & (0.70) & (0.26) & (0.76) & (0.49)\\
				
				F norm & 3.37 & 4.12 & 2.26 & 3.36 & 3.94 & \textbf{1.91}
				& 3.04 & 3.49 & \textbf{2.20} & 3.02 & 4.28 &  2.54 \\
				& (0.13) & (0.16) & (0.17) & (0.13) & (0.12) & (0.14)
				& (0.09) & (0.12) & (0.15) & (0.09) & (0.14) & (0.12) \\
				
				TPR & .9486 & .8794 & .6497 & \textbf{.9994} & .9760 & .8149 
				& .9936 & .9844 & .8376 & \textbf{.9998} & .9938 & .8671 \\		
				& (.0316) & (.0384) & (.0658) & (.0040) & (.0233) & (.0397)
				& (.0078) & (.0154) & (.0617) & (.0016) & (.0072) & (.0396) \\
				
				FPR & .1029 & .0442 & .0109 & .9983 & .1689 & \textbf{.0005}
				& .1029 & .0431 & \textbf{.0015} & .9988 & .1872 & .0033 \\
				& (.0150) & (.0077) & (.0029) & (.0055) & (.0066) & (.0003)
				& (.0161) & (.0093) & (.0009) & (.0029) & (.0066) & (.0011) \\
				
				Avg CPU time & 0.23 & 0.25 & 0.46 & 62.65 & 29.36 & 46.59
				& 0.19 & 0.20 & 0.33 & 73.36 & 30.74 & 45.82 \\
				
				\toprule
				& \multicolumn{6}{|c|}{Cliques positive} & \multicolumn{6}{|c|}{Cliques negative} \\
				nonzero pairs & \multicolumn{6}{|c|}{30/4950} & \multicolumn{6}{|c|}{30/4950} \\
				nonzero elements & \multicolumn{6}{|c|}{-0.45} & \multicolumn{6}{|c|}{0.75} \\
				$p=100, n=120$ & GL1 & GL2 & RGL & GSCAD & BGL & GHS & GL1 & GL2 & RGL & GSCAD & BGL & GHS \\
				\toprule
				
				Stein's loss & 4.60 & 7.14 & \textbf{1.26} & 4.57 & 42.69 & 1.89
				& 6.01 & 7.49 & \textbf{1.61} & 5.96 & 44.17 & 1.78 \\
				& (0.25) & (0.62) & (0.20) & (0.25) & (0.94) & (0.30)
				& (0.21) & (0.41) & (0.49) & (0.20) & (0.81) & (0.21) \\
				
				F norm & 2.82 & 3.85 & \textbf{1.63} & 2.80 & 3.83 & 1.98
				& 4.98 & 5.70 & \textbf{1.80} & 4.97 & 4.92 & 1.86 \\
				& (0.11) & (0.16) & (0.16) & (0.11) & (0.17) & (0.22)
				& (0.10) & (0.12) & (0.33) & (0.09) & (0.09) & (0.16) \\
				
				TPR & \textbf{1} & \textbf{1} & \textbf{1} & \textbf{1} & \textbf{1} & .9840
				& \textbf{1} & \textbf{1} & .9947 & \textbf{1} & \textbf{1} & \textbf{1} \\
				& (0) & (0) & (0) & (0) & (0) & (.0236)
				& (0) & (0) & (.0170) & (0) & (0) & (0) \\
				
				FPR & .0999 & .0287 & \textbf{.0003} & .9979 & .1580 & .0004
				& .0141 & .0413 & .0010 & .9939 & .1776 & \textbf{.0008} \\
				& (.0078) & (.0064) & (.0004) & (.0061) & (.0074) & (.0003)
				& (.0100) & (.0088) & (.0008) & (.0073) & (.0068) & (.0004) \\
				
				Avg CPU time & 0.16 & 0.17 & 0.62 & 4.60 & 32.63 & 37.96
				& 0.10 & 0.11 & 0.45 & 10.66 & 32.55 & 37.57 \\
				
				\bottomrule
		\end{tabular}}
		\addtolength{\tabcolsep}{3pt}
	\end{footnotesize}
\end{table}

\begin{table}[t]
	\centering
	\caption{Mean (sd) Stein's loss, Frobenius norm, true positive rates and false positive rates of precision matrix estimates over 50 data sets generated by multivariate normal distributions with precision matrix $\Omega_0$, where $p=200$ and $n=120$. The precision matrix is estimated by frequentist graphical lasso with penalized diagonal elements (GL1), frequentist graphical lasso with unpenalized diagonal elements (GL2), graphical SCAD (GSCAD), Bayesian graphical lasso (BGL), and graphical horseshoe (GHS). The best performer in each row is shown in bold. Average CPU time is in minutes.}
	
	\label{tab:3}
	\begin{footnotesize}
		\noindent\makebox[\textwidth]{%
			\begin{tabular}{l|rrrrr|rrrrr|}
				\toprule
				& \multicolumn{5}{|c|}{Random} & \multicolumn{5}{|c|}{Hubs} \\
				nonzero pairs & \multicolumn{5}{|c|}{29/19900} & \multicolumn{5}{|c|}{180/19900} \\
				nonzero elements & \multicolumn{5}{|c|}{$\sim -\mathrm{Unif}(0.2,1)$} & \multicolumn{5}{|c|}{0.25} \\
				$p=200, n=120$ & GL1 & GL2 & GSCAD & BGL & GHS & GL1 & GL2 & GSCAD & BGL & GHS \\
				\toprule   	
				
				Stein's loss & 10.06 & 15.80 & 9.96 & 116.61 & \textbf{2.94}
				& 12.49 & 15.12 & 12.40 & 122.87 & \textbf{11.67} \\
				& (0.40) & (0.99) & (0.39) & (1.69) & (0.34)
				& (0.45) & (0.80) & (0.42) & (1.35) & (0.76) \\
				
				F norm & 4.49 & 5.97 & 4.45 & 6.76 & \textbf{2.44} 
				& 4.61 & 5.27 & 4.59 & 7.10 & \textbf{3.74} \\
				& (0.14) & (0.17) & (0.14) & (0.18) & (0.16)
				& (0.10) & (0.15) & (0.08) & (0.16) & (0.14) \\
				
				TPR & .9476 & .8393 & \textbf{1} & .9855 & .8421
				& .9911 & .9773 & \textbf{1} & .9917 & .7754 \\
				& (.0370) & (.0301) & (0) & (.0232) & (.0369)
				& (.0065) & (.0132) & (0) & (.0060) & (.0323) \\
				
				FPR & .0514 & .0159 & .9951 & .1035 & \textbf{.0001}
				& .0657 & .0257 & .9997 & .1197 & \textbf{.0011} \\
				& (.0065) & (.0021) & (.0095) & (.0031) & ($<$.0001) 
				& (.0053) & (.0064) & (.0002) & (.0027) & (.0002) \\
				
				Avg CPU time & 0.03 & 0.04 & 562.13 & 934.53 & 1.44e+3
				& 0.03 & 0.03 & 266.00 & 750.57 & 767.55 \\
				
				\toprule
				& \multicolumn{5}{|c|}{Cliques positive} & \multicolumn{5}{|c|}{Cliques negative} \\
				nonzero pairs & \multicolumn{5}{|c|}{60/19900} & \multicolumn{5}{|c|}{60/19900} \\
				nonzero elements & \multicolumn{5}{|c|}{-0.45} & \multicolumn{5}{|c|}{0.75} \\
				$p=200, n=120$ & GL1 & GL2 & GSCAD & BGL & GHS & GL1 & GL2 & GSCAD & BGL & GHS \\
				\toprule
				
				Stein's loss & 11.59 & 17.79 & 11.53 & 124.93 & \textbf{4.09}
				& 14.56 & 18.12 & 14.50 & 126.44 & \textbf{3.77} \\
				& (0.37) & (0.84) & (0.34) & (1.69) & (0.39)
				& (0.29) & (0.80) & (0.29) & (1.45) & (0.37) \\
				
				F norm & 4.44 & 5.98 & 4.44 & 6.29 & \textbf{2.97}
				& 7.61 & 8.58 & 7.60 & 7.94 & \textbf{2.69} \\
				& (0.09) & (0.13) & (0.07) & (0.16) & (0.21)
				& (0.06) & (0.14) & (0.07) & (0.10) & (0.18) \\
				
				TPR & \textbf{1} & \textbf{1} & \textbf{1} & \textbf{1} & .9633
				& \textbf{1} & \textbf{1} & \textbf{1} & \textbf{1} & \textbf{1} \\
				& (0) & (0) & (0) & (0) & (.0226)
				& (0) & (0) & (0) & (0) & (0) \\
				
				FPR & .0663 & .0172 & .9969 & .0986 & \textbf{.0002}
				& .0636 & .0229 & .9919 & .1155 & \textbf{.0004} \\
				& (.0044) & (.0027) & (.0051) & (.0030) & (.0001)
				& (.0039) & (.0039) & (.0072) & (.0027) & (.0001) \\
				
				Avg CPU time & 0.02 & 0.03 & 192.46 & 1.64e+03 & 1.70e+03
				& 0.07 & 0.04 & 466.89 & 1.06e+03 & 847.92 \\
				
				\bottomrule
		\end{tabular}}
	\end{footnotesize}
\end{table}

	\subsection{Variable Selection}
	\label{sec:var_selection}

	\citet{van2017uncertainty} studied the coverage properties of marginal credible intervals under the horseshoe prior, for a sparse normal means problem. They found that the model selection procedure using credible intervals under the horseshoe prior is conservative. That is, few zero parameters in the model are falsely selected,  but some of the signals are not selected. In simulations, they also discovered that the lengths of credible intervals under the horseshoe prior adapt to the signal size. In other words, parameters with larger nonzero means have wider credible intervals. In order to reduce false negatives due to wide credible intervals for large signals, we use the $50\%$ credible interval for variable selection. By the conservative property of the procedure, false positives would be controlled under this criterion. This choice also agrees with the median probability model suggested by \citet{barbieri2004optimal}.
	
\begin{figure*}[t]
	\centering
	\begin{subfigure}[t]{0.24\textwidth}
		\centering
		\adjincludegraphics[width=\textwidth,trim={{0.12\width} {0.05\height} {0.16\width} {0\height}},clip]{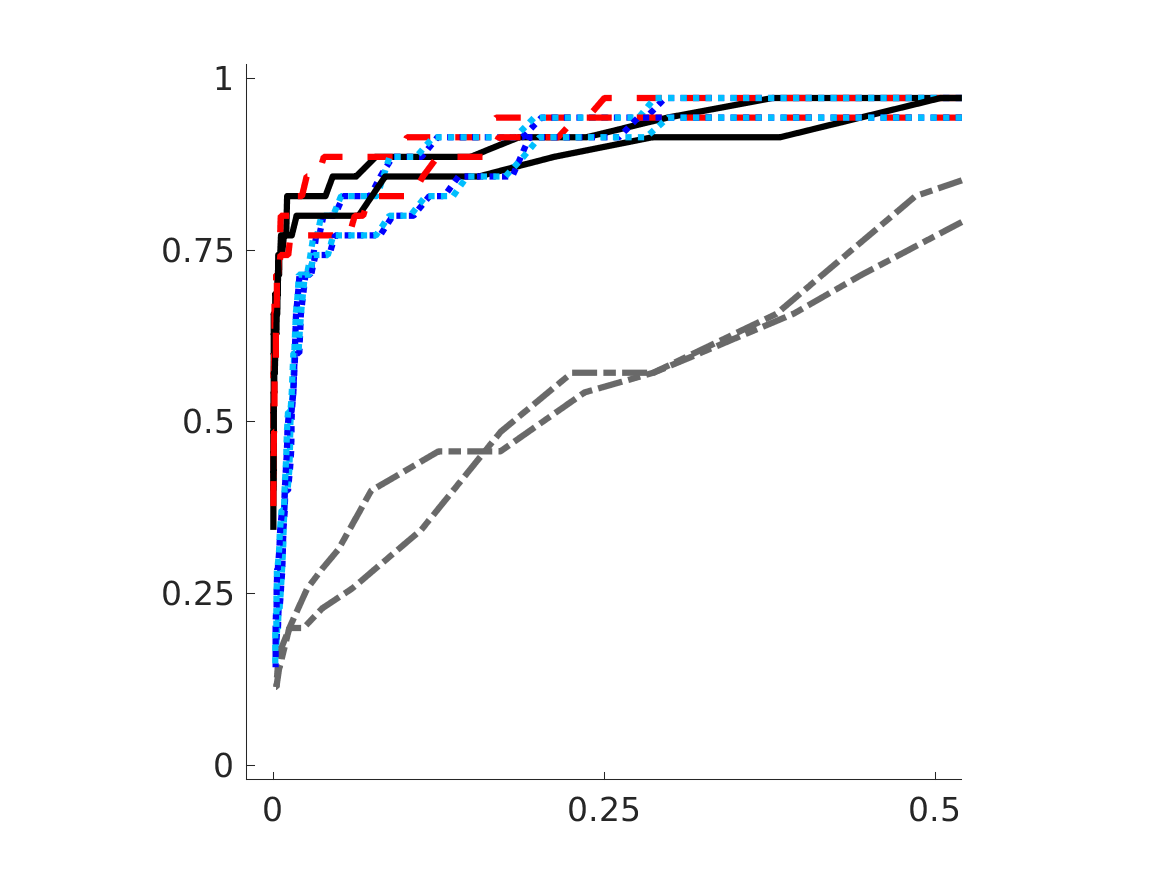}
		\caption[ROC]%
		{{\footnotesize Random structure}}
		\label{fig:ROC_p100random}
	\end{subfigure}
	\
	\begin{subfigure}[t]{0.24\textwidth}
		\centering
		\adjincludegraphics[width=\textwidth,trim={{0.12\width} {0.05\height} {0.16\width} {0\height}},clip]{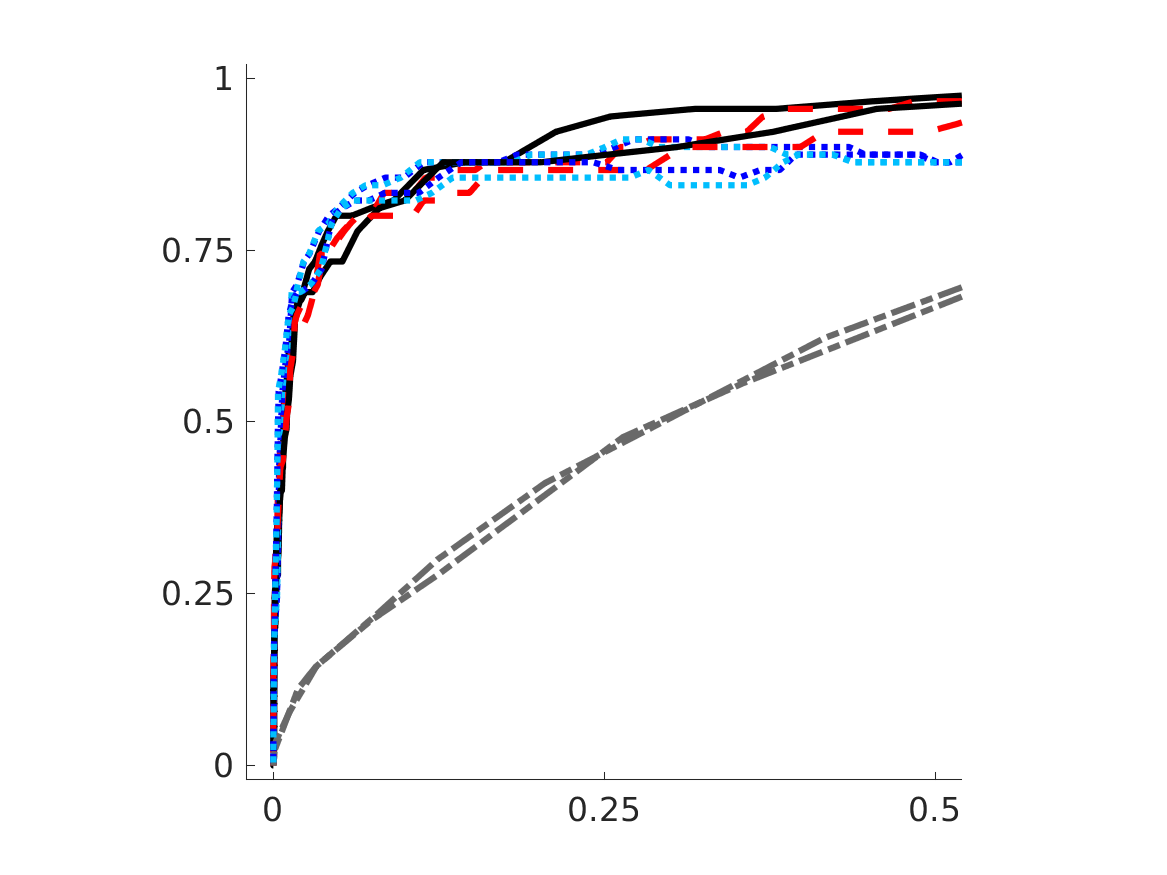}
		\caption[]%
		{{\footnotesize Hubs structure}}
		\label{fig:ROC_p100hubs}
	\end{subfigure}
	\
	\begin{subfigure}[t]{0.24\textwidth}
		\centering
		\adjincludegraphics[width=\textwidth,trim={{0.12\width} {0.05\height} {0.16\width} {0\height}},clip]{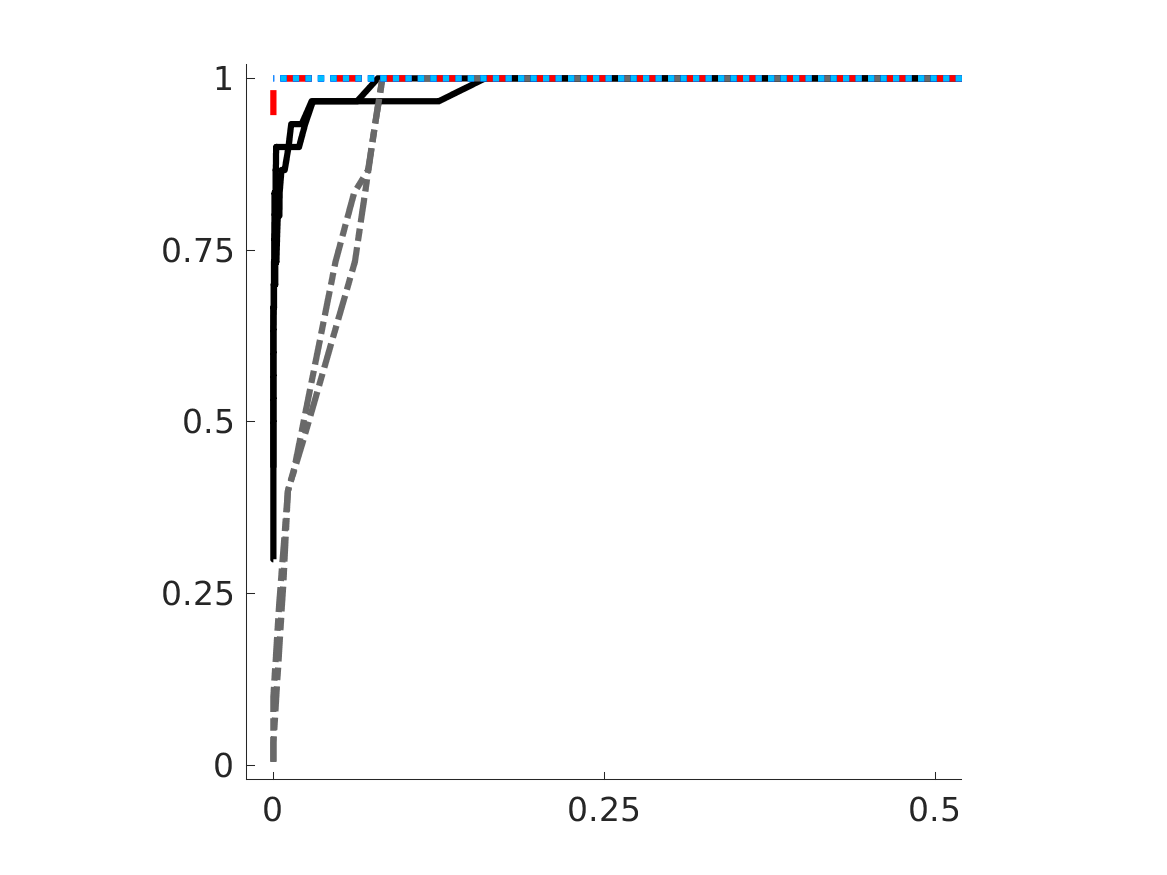}
		\caption[]%
		{{\footnotesize Cliques structure, positive}}
		\label{fig:ROC_p100cliquespos45}
	\end{subfigure}
	\
	\begin{subfigure}[t]{0.24\textwidth}
		\centering
		\adjincludegraphics[width=\textwidth,trim={{0.12\width} {0.05\height} {0.16\width} {0\height}},clip]{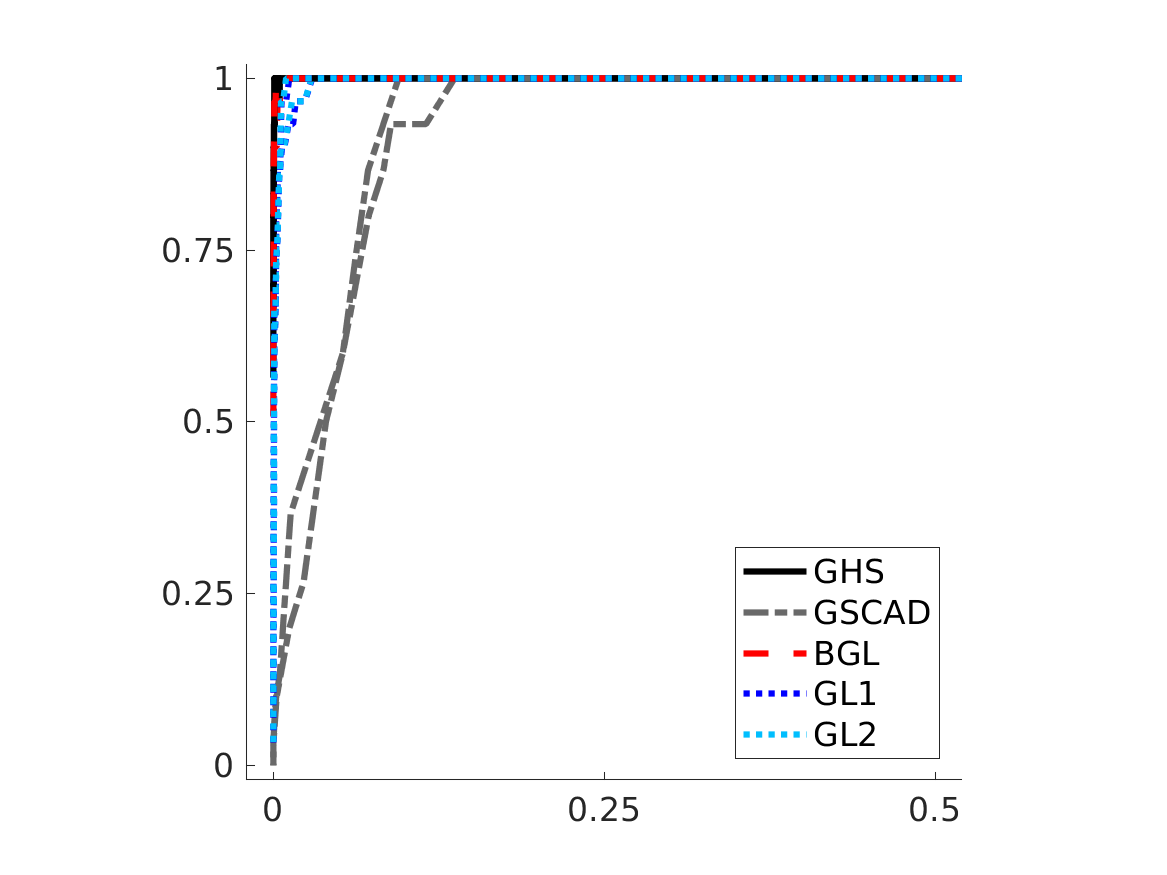}
		\caption[]%
		{{\footnotesize Cliques structure, negative}}
		\label{fig:ROC_p100cliquesneg75}
	\end{subfigure}
	\caption{Receiver operating characteristic (ROC) curves of estimates by frequentist graphical lasso with penalized diagonal elements (GL1), frequentist graphical lasso with unpenalized diagonal elements (GL2), graphical SCAD (GSCAD), Bayesian graphical lasso (BGL), and graphical horseshoe (GHS), for precision matrix with random structure, hubs structure, cliques structure with positive partial correlations, and cliques structure with negative partial correlations. $p=100$ and $n=50$. The true positive rate is shown on the y-axis, and the false positive rate is shown on the x-axis. ROC curves of two representative data sets in simulations.}
	\label{fig:ROC}
\end{figure*}

	True and false positive rates are reported in Table~\ref{tab:1}, \ref{tab:2} and \ref{tab:3}. True positive rates under the graphical horseshoe prior are indeed lower when $p=100$ and $n=50$. However,  the true positive rate for the graphical horseshoe improves greatly when $n=120$. The graphical horseshoe also has lower false positive rates than the other regularization methods. Figure~\ref{fig:ROC} shows the ROC curves, plotting true positive rate against false positive rate for variable selection results, when $p=100$ and $n=50$. To avoid overlapping curves, ROC curves of two representative data sets were plotted in each case. The ROC curves for the graphical lasso and graphical SCAD are generated by estimating the precision matrix with a sequence of various tuning parameters. The ROC curves for the graphical horseshoe and Bayesian graphical lasso are generated by varying the length of posterior credible intervals, from $1\%$ to $99\%$. Except for the graphical SCAD, which always performs worse in variable selection, the other methods have similar ROC curves. For the random and cliques structured matrix with negative partial correlations, the ROC curve of the graphical horseshoe is slightly closer to the y-axis. Although the difference is minute in terms of the false positive rate, such a difference could greatly increase precision, the rate of true positives among all discoveries, in a sparse model. When most parameters are zero, a little increase in false discovery rate greatly increases the number of false discoveries and decreases precision. In our simulations, the precision for the graphical horseshoe is almost always higher than $0.85$, while the precision for other regularization methods is usually less than $0.3$, making the variable selection results not very useful in applications. Additional numerical results on precision of the estimates in simulations can be found in Tables~S.1, S.2 and S.3 of the Supplementary Material.
	
	Finally, it is worth noting that there need not to be a single variable selection result by a Bayesian model. In applications, researchers can obtain posterior samples from the graphical horseshoe or Bayesian graphical lasso, and gradually change the length of credible intervals for variable selection to have a sequence of results following the ROC curve. Such a procedure allows the researcher to start from a low false positive rate and moderate true positive rate, and gradually increase the true positive rate while having some control on precision.

\section{Analysis of Human Gene Expression Data}
\label{sec:real data}

We analyze the expression of 100 genes in 60 unrelated individuals of Northern and Western European ancestry from Utah (CEU). A description of the data set can be found in \citet{bhadra2013joint}. For this analysis, we assume that the gene expressions of the individuals in this data set are identically distributed with a multivariate normal distribution. We analyze centered gene expressions using graphical horseshoe, graphical lasso with penalized diagonal elements, graphical SCAD, and Bayesian graphical lasso. Tuning parameters in graphical lasso and graphical SCAD are selected by five--fold cross validation, using log likelihood. For graphical lasso and graphical SCAD, the existence of association between a pair of genes in terms of expression is determined by whether the corresponding element in the precision matrix is estimated to be zero. For the graphical horseshoe and Bayesian graphical lasso, we used whether zero is included in the $50\%$ posterior credible interval.

\begin{figure}[H]
	\centering
	\begin{subfigure}[b]{0.325\textwidth}
		\centering
		\adjincludegraphics[width=\textwidth,trim={{0.15\width} {0.1\height} {0.15\width} {0.1\height}},clip]{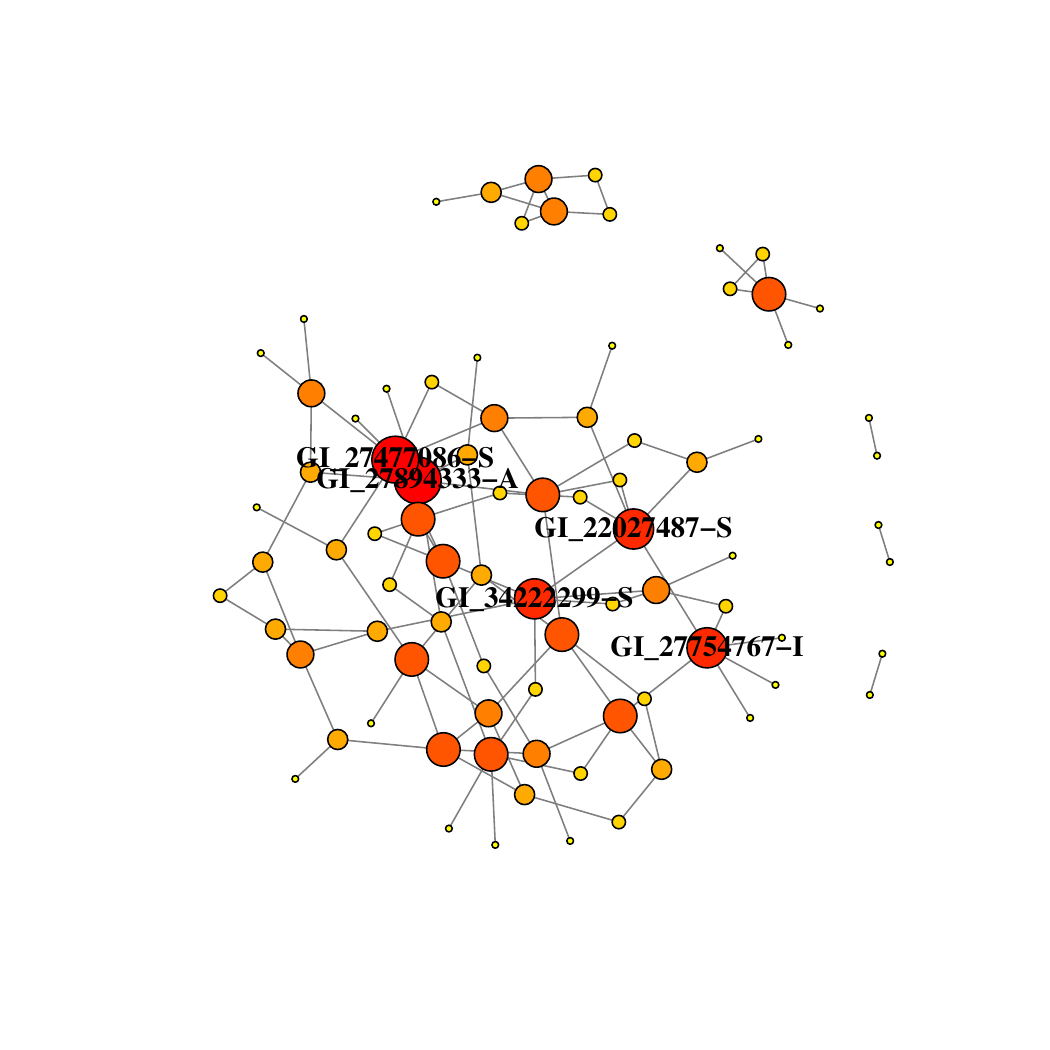}	
		\caption[geneExpression]%
		{{\footnotesize Estimated graph by GHS}}
		\label{fig:geneexp_GHS}
	\end{subfigure}
	\begin{subfigure}[b]{0.325\textwidth}
		\centering
		\adjincludegraphics[width=\textwidth,trim={{0.15\width} {0.1\height} {0.15\width} {0.1\height}},clip]{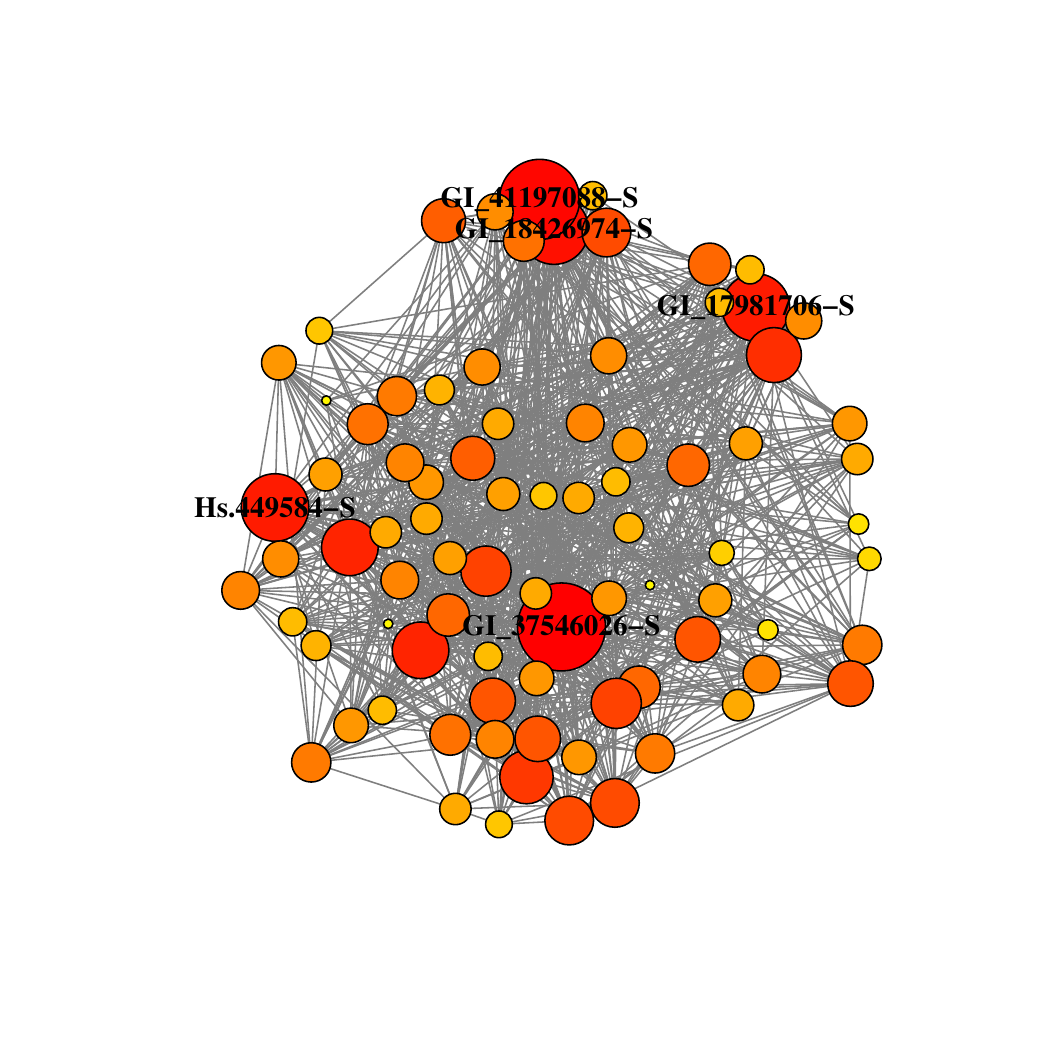}	
		\caption[geneExpression]%
		{{\footnotesize Estimated graph by GL}}
		\label{fig:geneexp_GL}
	\end{subfigure}
	\begin{subfigure}[b]{0.325\textwidth}
		\centering
		\adjincludegraphics[width=\textwidth,trim={{0.15\width} {0.1\height} {0.15\width} {0.1\height}},clip]{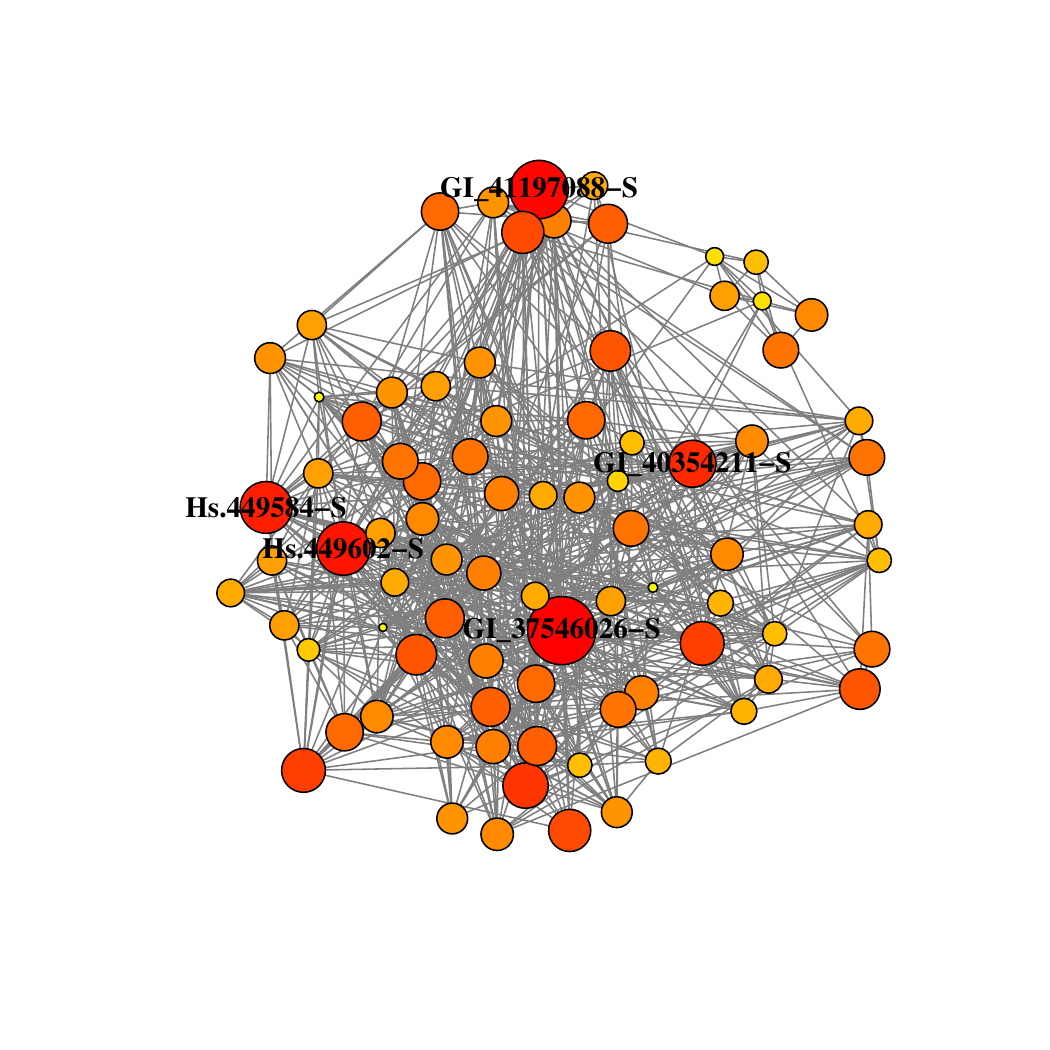}	
		\caption[geneExpression]%
		{{\footnotesize Estimated graph by BGL}}
		\label{fig:geneexp_BGL}
	\end{subfigure}
	\caption{The inferred graph for the CEU data, by graphical horseshoe (GHS), frequentist graphical lasso with penalized diagonal elements (GL), and Bayesian graphical lasso (BGL) estimates. Genes that are conditionally independent of all the others are not shown. Size of node is proportional to degree within each graph, the positions of nodes are comparable across graphs.}
	\label{fig:geneexp}
\end{figure}

The inferred graph by graphical horseshoe, graphical lasso and Bayesian graphical lasso are shown in Figure~\ref{fig:geneexp}. The graphical horseshoe estimate has 83 vertices and 109 edges. The inferred graph has 100 vertices and 1135 edges by graphical lasso estimate, and 100 vertices and 976 edges by Bayesian graphical lasso estimate. None of the graphical SCAD estimated elements in the precision matrix is zero, so the inferred graph by graphical SCAD estimate has 100 vertices and 4950 edges. The graphs by graphical lasso and Bayesian graphical lasso show similar clusters, where every gene expression is associated with at least one other gene expression, and the major clusters are densely connected as well. On the other hand, the graphical horseshoe estimate shows unconnected and much sparser clusters of gene expressions. 
Our resulting network using this human gene expression data can be compared with that in \citet{bhadra2013joint}, who used the same data set in a regression setting (as opposed to the zero mean setting used by us), where the gene expressions were regressed on SNPs and the resulting network on the residual terms was plotted. Comparison of these two networks should provide an insight into which edges are ``robust'' to the effect of being conditioned upon the SNPs.

\section{Conclusions}\label{sec:conc}
\label{sec:conclusion}

The problem of precision matrix estimation in a multivariate Gaussian model poses a challenge in high-dimensional data analysis. In this paper, we proposed the graphical horseshoe estimator with easy implementation by a full Gibbs sampler. By using a prior with high density near the origin and a Cauchy-distributed local shrinkage parameter on each dimension, the graphical horseshoe model generates estimates close to the true distribution in Kullback--Leibler divergence and with small bias for nonzero elements. Simulations confirm that the graphical horseshoe outperforms alternative methods in various situations.

We have shown when the Kullback--Leibler divergence is under consideration, all methods eventually fail in high dimensions. In addition, the difference between sample size and feature size also affects bias. This implies that efforts should be spent on variable screening prior to analysis in order to bring the feature space to a manageable size. 
Although some properties of variable selection by the horseshoe prior in sparse normal means problem are known, theoretical understanding of true and false discoveries under the graphical horseshoe prior are still lacking. It will also be interesting to compare the graphical horseshoe to some recently proposed methods in graphical model estimation, for instance, the spike-and-slab lasso \citep{deshpande2017simultaneous}. Use of other priors exhibiting properties similar to the horseshoe, such as the horseshoe+ \citep{bhadra2017horseshoe+} or the Dirichlet--Laplace \citep{bhattacharya2015dirichlet} should also be explored.

\section*{Supplementary Material}

The Supplementary Material contains proofs of theorems, MCMC convergence diagnostics and additional simulation results.

\section*{Acknowledgements}

We thank the AE and two anonymous referees for many constructive suggestions. The research of Bhadra is partially supported by Grant No. DMS-1613063 by the US National Science Foundation.

\bibliographystyle{biom}
\bibliography{GHS}

\clearpage\pagebreak\newpage
\thispagestyle{empty}
\begin{center}
	{\LARGE{\bf Supplementary Material to\\ {\it The Graphical Horseshoe Estimator for Inverse Covariance Matrices}}}
\end{center}

\vskip 2cm
\baselineskip=15pt

\begin{center}
	\vspace{-1cm}
	Yunfan Li\\
	Department of Statistics, Purdue University, 250 N. University Street, West Lafayette, IN 47907, USA.\\
	li896@purdue.edu\\
	\hskip 5mm \\
	Bruce A.~Craig\\
	Department of Statistics, Purdue University, 250 N. University Street, West Lafayette, IN 47907, USA.\\
	bacraig@purdue.edu\\
	\hskip 5mm \\
	Anindya Bhadra\\
	Department of Statistics, Purdue University, 250 N. University Street, West Lafayette, IN 47907, USA.\\
	bhadra@purdue.edu\\
\end{center}

\setcounter{equation}{0}
\setcounter{page}{0}
\setcounter{table}{0}
\setcounter{section}{0}
\setcounter{subsection}{0}
\setcounter{figure}{0}
\renewcommand{\theequation}{S.\arabic{equation}}
\renewcommand{\thesection}{S.\arabic{section}}
\renewcommand{\thesubsection}{S.\arabic{subsection}}
\renewcommand{\thepage}{S.\arabic{page}}
\renewcommand{\thetable}{S.\arabic{table}}
\renewcommand{\thefigure}{S.\arabic{figure}}

\newpage

\subsection{Proof of Theorem~3.2}
\label{app:bound}

\begin{spacing}{1.5}
	
	We claim that an Euclidean cube of $p^2$ dimensions with $(\omega_{ij0},\omega_{ij0}+\sqrt{\epsilon}/2Mp)$ on each dimension lies inside $A_{\epsilon}$ and an Euclidean cube with $p^2$ dimensions with $(\omega_{ij0}-2\sqrt{\epsilon}/Mp,\omega_{ij0}+2\sqrt{\epsilon}/Mp)$ on each dimension contains $A_{\epsilon}$. The proof is as following.
	
	$D(p_{\Omega_0} ||  p_{\Omega})=\frac{1}{2}\{\text{log}|\Omega^{-1}\Omega_0|+\text{tr}(\Omega\Omega_0^{-1})-p\}$. Take $\Omega=\Omega_0+(\delta/Mp)\mathbb{1}$ where $\mathbb{1}$ is a matrix with all elements equal to 1, then
	\begin{align*}
	\textnormal{tr}(\Omega_0^{-1}\Omega)-p=& \textnormal{tr}(\Omega_0^{-1}(\Omega_0+(\delta/Mp)\mathbb{1}))-\textnormal{tr}(\Omega_0^{-1}\Omega_0) \\
	=& \textnormal{tr}(\Omega_0^{-1}(\delta/Mp)\mathbb{1}) \\
	=& \sum_{i,j}\sigma_{ij0}*(\delta/Mp)=\delta, \\
	\textnormal{log}|\Omega^{-1}\Omega_0|=& \textnormal{log}|(\Omega_0+(\delta/Mp)\mathbb{1})^{-1}\Omega_0| \\
	=& -\textnormal{log}|\Omega_0^{-1}(\Omega_0+(\delta/Mp)\mathbb{1})| \\
	=& -\textnormal{log}|I+\Omega_0^{-1}(\delta/Mp)\mathbb{1})| \\
	=& -\textnormal{log} \prod_{i=1}^p (1+\lambda_i) \\
	=& -\sum_{i=1}^p \textnormal{log}(1+\lambda_i),
	\end{align*}
	where $\lambda_i$ are the eigenvalues of the matrix $\Omega_0^{-1}(\delta/Mp)\mathbb{1}$. The matrix $\Omega_0^{-1}(\delta/Mp)\mathbb{1}$ has column rank $1$, and its only non-zero eigenvalue is equal to $\sum_{i,j}\sigma_{ij0}*(\delta/Mp)=\delta$. Therefore $D(p_{\Omega_0} || p_{\Omega})=\delta - \text{log}(1+\delta)$. The function $x-\text{log}(1+x)$ has expansion $x^2/2+O(x^3)$ at $x=0$. Take $\delta=\sqrt\epsilon/2$ and $2\sqrt\epsilon$, and it can be verified that the claim at the beginning of this proof is true when $\sqrt\epsilon \to 0$.
	
	Now that we find cubes that lies in and contains $A_{\epsilon}$, we can bound $\nu(A_{\epsilon})$ by the product of prior measures on each dimension of these cubes. For any prior $p(\omega_{ij})$ satisfying the conditions stated in the part (2) of the theorem, $\int_{\omega_{ij0}-\sqrt\epsilon/Mp}^{\omega_{ij0}+\sqrt\epsilon/Mp} p(\omega_{ij}) d\omega_{ij} \propto \frac{\sqrt\epsilon}{Mp}$ since the density is bounded above. The horseshoe prior also satisfies these conditions when $\omega_{ij0}\neq 0$ \citep{carvalho2010horseshoe}, so the same formula holds for graphical horseshoe prior when $\omega_{ij0}\neq 0$. Taking $\epsilon=1/n$ and summing over $p^2$ dimensions completes the proof of Part (2) of Theorem 3.2.
	
	By Theorem 1 in \citet{carvalho2010horseshoe}, the horseshoe prior has tight bounds when $\tau=1$. Using these bounds, $K/2 \int_0^{\sqrt\epsilon/Mp} \text{log}(1+4/\omega_{ij}^2) d\omega_{ij} < \int_0^{\sqrt\epsilon/Mp} p(\omega_{ij}) d\omega_{ij}$ when $\omega_{ij0}=0$, $K=1/\sqrt{2\pi^3}$. Using the variable change in the proof of Theorem 4 in \citet{carvalho2010horseshoe}, let $u=4/\omega_{ij}^2$, then integrate by parts
	\begin{align*}
	& \int_0^{\sqrt\epsilon/Mp} \text{log}(1+4/\omega_{ij}^2) d\omega_{ij} \\
	=& \int_{4M^2p^2/\epsilon}^\infty \text{log}(1+u)u^{-3/2}\text{d}u \\
	=& \frac{2\sqrt{\epsilon}}{Mp} \text{log} \left( 1+\frac{4M^2p^2}{\epsilon} \right)+4 \left(\frac{\pi}{2}-\text{arctan}\sqrt{\frac{4M^2p^2}{\epsilon}} \right).
	\end{align*}
	After some algebra and taking $\epsilon=1/n$, the final expression is $\frac{2}{M\sqrt{n}p}\text{log}(1+4M^2 n p^2)+\frac{2}{M\sqrt{n}p}-O\{(\frac{1}{4M^2np^2})^{3/2}\} > \frac{4}{M\sqrt{n}p}\text{log}(2M\sqrt{n}p)$. Having fixed values of $\tau$ other than $1$ does not change the rate of this integration with respect to $\sqrt\epsilon/Mp$. Part (1) of Theorem~3.2 is derived.
	
\end{spacing}

\subsection{Proof of Theorem 4.1}
\label{app:bias}

\begin{spacing}{1.5}
	
	First, consider the posterior mean estimate under the graphical horseshoe prior. It is obvious that $\hat{\omega}_{pj}'\, | \, Y_{(-p)} \sim \text{Normal}(\omega_{pj0}', 1)$ and $\hat{\omega}_{pj}'^2\, | \, Y_{(-p)} \sim \text{Noncentral} \, \chi^2 (1, \omega_{pj0}'^2)$. From the horseshoe prior, $\omega_{pj} \sim \text{Normal}(0, \lambda_{pj}^2\tau^2)$ and $\omega_{pj}' \sim \text{Normal}(0, \lambda_{pj}^2\tau^2\omega_{pp}^{-1}m^{-1})$ where $m=\{(Y_{(-p)}'Y_{(-p)})^{-1}_{jj}\}^{-1}$. We use $\omega_{pp}$ and $\omega_{pp0}$ interchangeably in the proof since all the diagonal elements are assumed known. Then
	\begin{equation*}
	\hat{\omega}_{pj}'\, |\, Y_{(-p)},\lambda_{pj}^2,\tau^2, \sim \text{Normal}(0, 1+\lambda_{pj}^2\tau^2\omega_{pp}^{-1}m^{-1}).
	\end{equation*}
	To get the marginal distribution of $\hat{\omega}_{pj}'$, integrate the local shrinkage parameter $\lambda_{pj}$,
	\begin{equation*}
	m(\hat{\omega}_{pj}'^2) \propto \int_0^{\infty} (2\pi)^{-1/2} \pi^{-1} (1+\lambda_{pj}^2\tau^2\omega_{pp}^{-1}m^{-1})^{-1/2} 
	\text{exp} \left\{ -\frac{\hat{\omega}_{pj}'^2}{2(1+\lambda_{pj}^2\tau^2\omega_{pp}^{-1}m^{-1})} \right\}
	\frac{1}{1+\lambda_{pj}^2} \text{d}\lambda_{pj}.
	\end{equation*}
	Let $Z_{pj}=1/(1+\lambda_{pj}^2\tau^2\omega_{pp}^{-1}m^{-1})$ so that the Jaobian is $\frac{\partial \lambda_{pj}}{\partial Z_{pj}}=-\frac{1}{2} \left\{ \frac{1}{\omega_{pp}^{-1}m^{-1}\tau^2}(\frac{1}{Z_{pj}}-1) \right\}^{-1/2} \frac{Z_{pj}^{-2}}{\omega_{pp}^{-1}m^{-1}\tau^2}$, then 
	\begin{equation*}
	m(\hat{\omega}_{pj}'^2) \propto \int_0^1 \text{exp} \left( -\frac{Z_{pj}\hat{\omega}_{pj}'^2}{2} \right) (1-Z_{pj})^{-1/2} \left\{ \frac{1}{\omega_{pp}^{-1}m^{-1}Z_{pj}^2}+(1-\frac{1}{\omega_{pp}^{-1}m^{-1}\tau^2})Z_{pj} \right\}^{-1} \text{d}Z_{pj}.
	\end{equation*}
	This expression differs only by a scale $\omega_{pp}^{-1}m^{-1}$ from expressions leading to Proposition 4.1 in \citet{bhadra2016prediction}. Using proof of Theorem 4.1 in \citet{bhadra2016prediction} and Theorem 2 in \citet{carvalho2010horseshoe}, the posterior mean under the horseshoe prior is $\mathrm{E}(\omega_{pj}'\, | \, Y, \tau)=(1-\mathrm{E}(Z_{pj}))\hat{\omega}_{pj}'$, where $Z_{pj} \sim \text{CCH}(1,1/2,1,\hat{\omega}_{pj}'^2/2,1,1/\omega_{pp}^{-1}m^{-1}\tau^2)$. Let $\theta_{pj}=1/(\omega_{pp}^{-1}m^{-1}\tau^2)$, then an upper bound for $\mathrm{E}(Z_{pj})$ is $4(C_1+C_2)\theta_{pj}(1+\hat{\omega}_{pj}'^2/2)/\hat{\omega}_{pj}'^4$ when $\hat{\omega}_{pj}'^2/2>1$, where $C_1=1-2e$ and $C_2=\Gamma(1/2)\Gamma(2)/\Gamma(2.5)$ by Theorem 4.2 in \citet{bhadra2016prediction}. Consequently, $\mathrm{E}(Z_{pj})$ is $O(1/\hat{\omega}_{pj}'^2)$ when $\hat{\omega}_{pj}' \to \infty$, completing the proof of Part (1).
	
	Now consider the posterior mean estimate under the double-exponential prior in Part (2). Since double-exponential distribution is a scale mixture of normals \citep{park2008bayesian}, the posterior mean estimate has expression $\mathrm{E}(\omega_{pj}'\, | \, Y)_{lasso}= \hat{\omega}_{pj}'+ \frac{\mathrm{d}}{\mathrm{d}\hat{\omega}_{pj}'}\mathrm{log} \, m_{lasso}(\hat{\omega}_{pj}')$ by Theorem 2 in \citet{carvalho2010horseshoe}. Equation (5) in \citet{carvalho2009handling} states that $lim_{|\hat{\omega}_{pj}'|\to\infty} \frac{\mathrm{d}}{\mathrm{d}\hat{\omega}_{pj}'}\mathrm{log} \, m_{lasso}(\hat{\omega}_{pj}')=\pm a$, where $a$ varies inversely with the global shrinkage parameter in the prior. The proof of this statement is in \citet{pericchi1992exact}.
	
	Now consider the condition that $\hat{\omega}_{pj}'^2$ is large. $(Y_{(-p)}'Y_{(-p)})^{-1}$ follows an inverse Wishart distribution with scale matrix $\Sigma_{(-p)}^{-1}$ and $n$ degrees of freedom, where $\Sigma_{(-p)}$ is the covariance matrix without the $p$th column and $p$th row. Consequently, $(Y_{(-p)}'Y_{(-p)})^{-1}_{jj}$ follows a one-dimensional inverse Wishart distribution with scale $\Sigma_{(-p)jj}^{-1}$ and $n-p+2$ degrees of freedom, and its inverse $\{(Y_{(-p)}'Y_{(-p)})^{-1}_{jj}\}^{-1}$ follows a Wishart distribution with scale $\{\Sigma_{(-p)jj}^{-1}\}^{-1}$ and $n-p+2$ degrees of freedom, or equivalently a gamma distribution with shape parameter $(n-p+2)/2$ and scale parameter $2\{\Sigma_{(-p)jj}^{-1}\}^{-1}$. By matrix inversion in blocked form, $\Sigma_{(-p)}^{-1}=\Omega_{(-p)}-\bm{\omega}_{(-p)p}\bm{\omega}_{p(-p)}/\omega_{pp}$, so that the scale parameter of the gamma distribution is $2(\omega_{jj0}-\omega_{jp0}^2/\omega_{pp0})^{-1}$, as claimed in Part (3).
	
\end{spacing}

\subsection{MCMC Convergence Diagnostics} \label{sec:sup_MC}
\label{sec:mcmc}

\begin{figure}[h]
	\centering
	\begin{subfigure}{.31\textwidth}
		\centering
		\adjincludegraphics[width=\textwidth,trim={{0.02\width} {0\height} {0.02\width} {0\height}},clip]{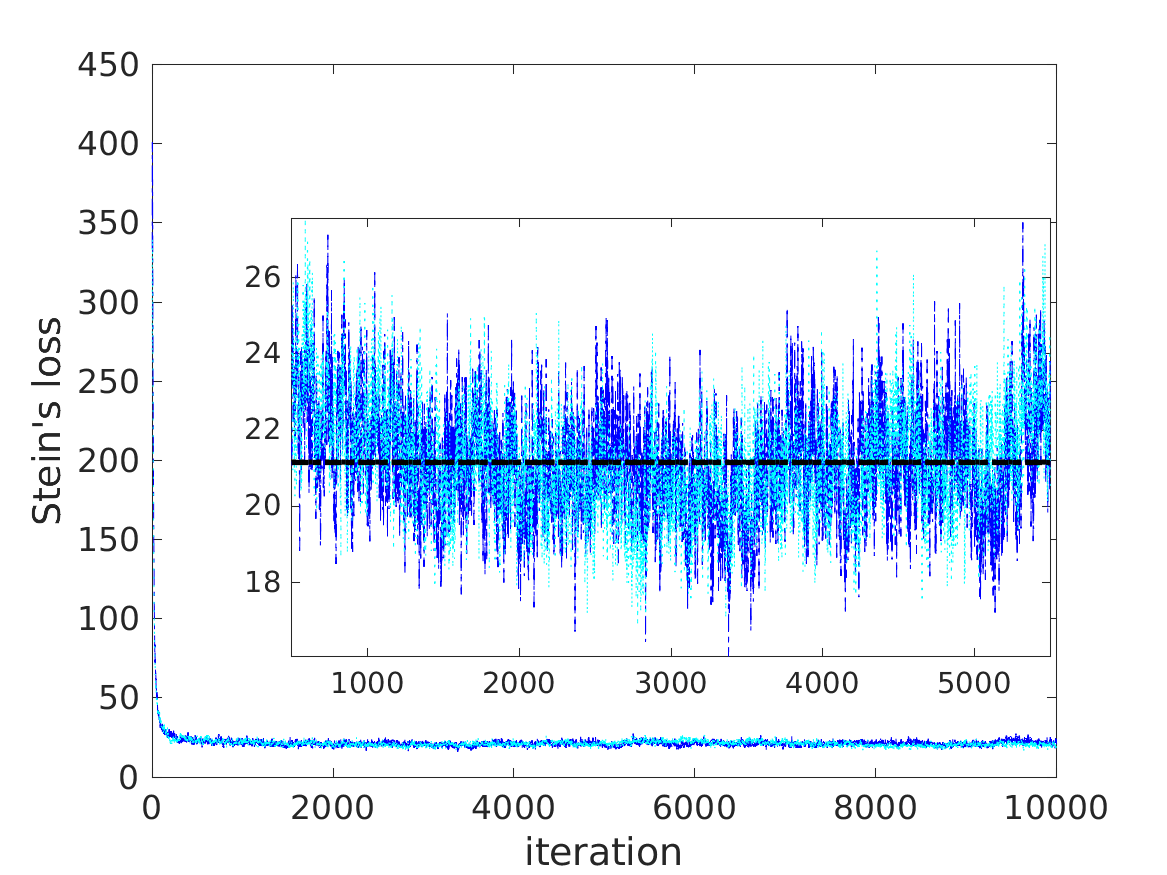}
		\caption{$p=100$, $n=50$}
		\label{fig:Sloss_sub1}
	\end{subfigure}%
	\begin{subfigure}{.31\textwidth}
		\centering
		\adjincludegraphics[width=\textwidth,trim={{0.02\width} {0\height} {0.02\width} {0\height}},clip]{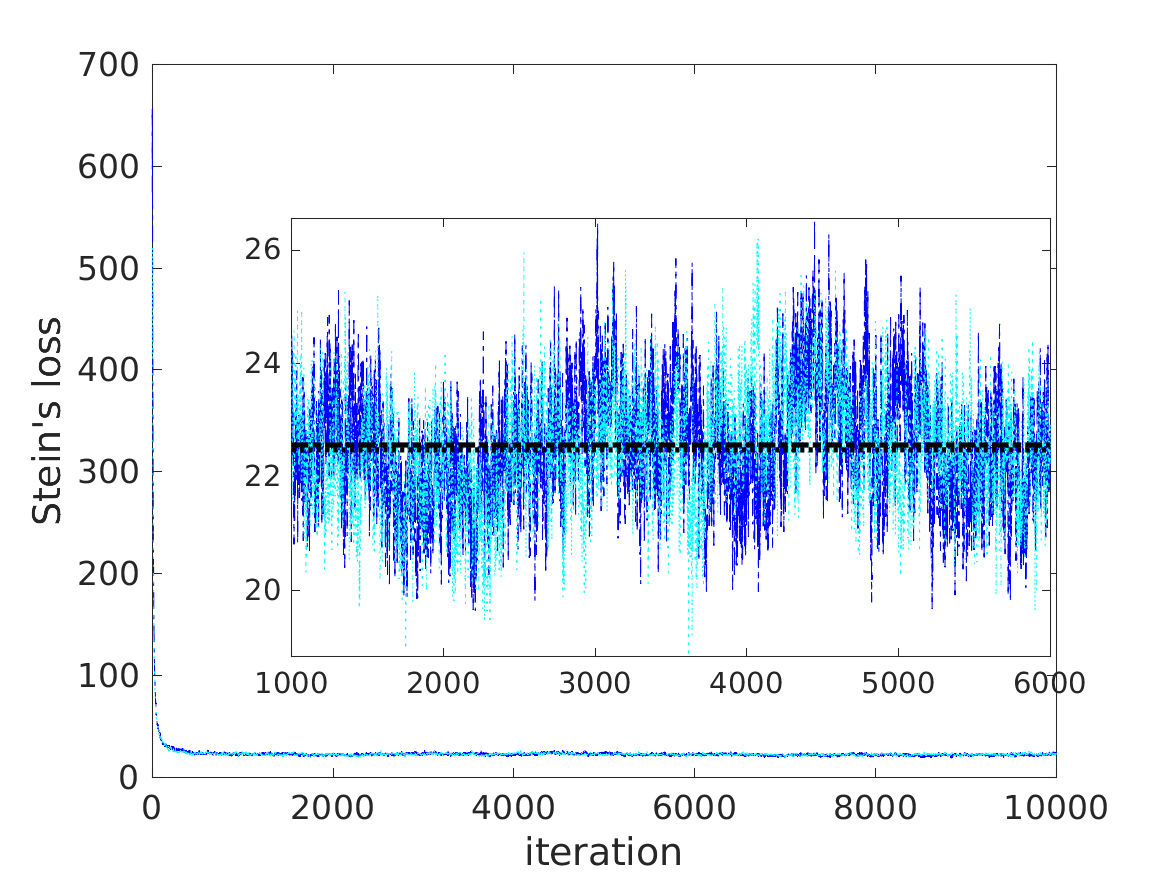}
		\caption{$p=200$, $n=120$}
		\label{fig:Sloss_sub2}
	\end{subfigure}
	\begin{subfigure}{.31\textwidth}
		\centering
		\adjincludegraphics[width=\textwidth,trim={{0.02\width} {0\height} {0.02\width} {0\height}},clip]{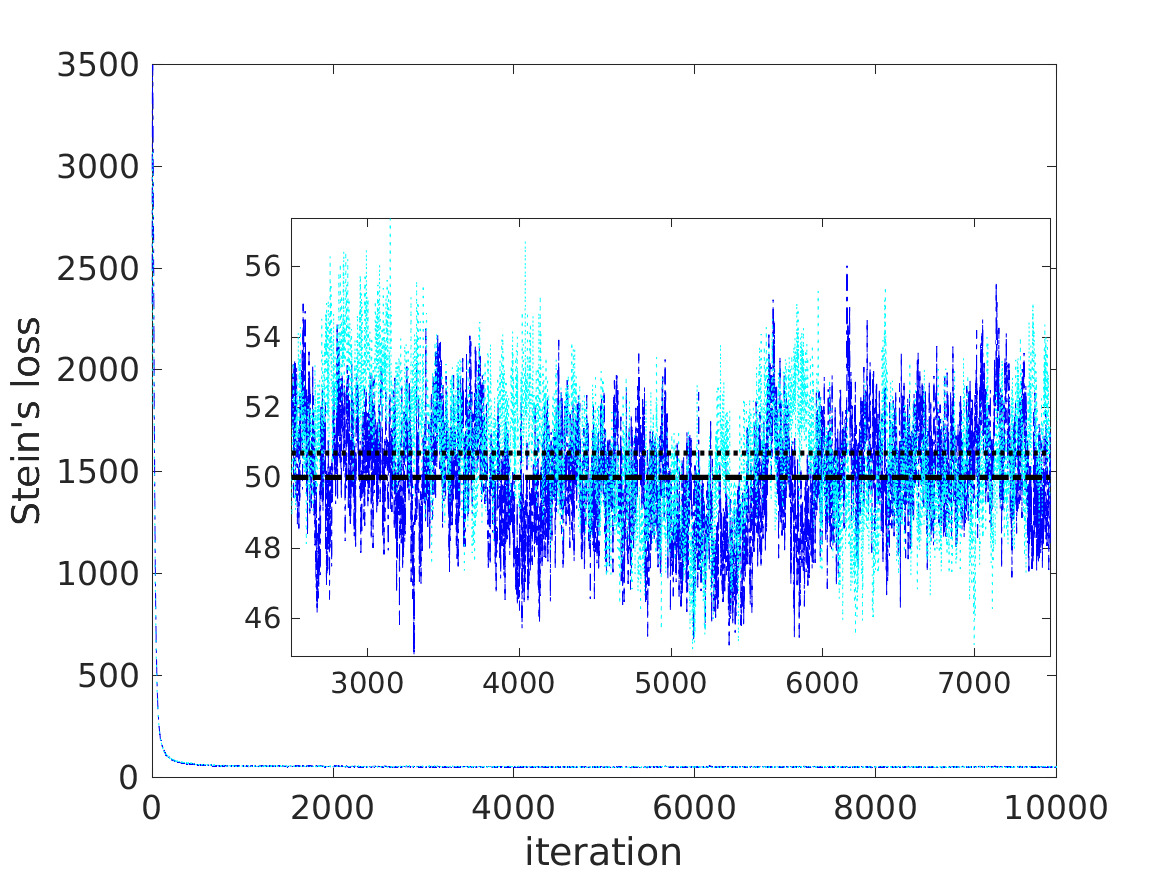}
		\caption{$p=400$, $n=120$}
		\label{fig:Sloss_sub3}
	\end{subfigure}
	\caption{Stein's loss of the sampled $\Omega$ at each iteration using Algorithm~1 for graphical horseshoe, under (a) hubs structure, $p=100$, $n=50$, and (b) hubs structure, $p=200$, $n=120$, (c) hubs structure, $p=400$, $n=120$. The first data set in the corresponding simulations are used. The dashed line and dotted line show Stein's loss of samples from two chains with different starting values, a $p \times p$ identity matrix and a random $p \times p$ positive definite symmetric matrix. The enlarged area show Stein's loss in a shorter range of iterations, and the horizontal lines show the average Stein's loss of iterations within that range.}
	\label{fig:Sloss}
\end{figure}
We evaluate the convergence and mixing of the proposed graphical horseshoe Gibbs sampler by plotting Stein's loss of sampled $\Omega$ across iterations (i.e., a trace plot), using different starting values for each chain. It is shown that when $p=100$ and $p=200$, MCMC samples using Algorithm 1 converge within 500 iterations, and mix reasonably well. When $p=400$, the algorithm takes longer to converge. In the simulations, we use more burn-in samples when the dimension is higher. We use $500$ burn-in samples when $p=100$; $1000$ burn-in samples when $p=200$; and $2500$ burn-in samples when $p=400$. We use $5000$ iterations in all cases, for both graphical horseshoe and Bayesian graphical lasso. Figure~\ref{fig:Sloss} shows the plots used for MCMC diagnostics. Formal tests such as Gelman--Rubin diagnostics could be carried out using the MCMC output, if desired.

\subsection{Additional Simulation Results}

We provide additional simulation results in this section. The purpose is two-fold:

\begin{enumerate}
	\item Tables~\ref{stab:1}, \ref{stab:2} and~\ref{stab:3} provide estimates of sensitivity, specificity, precision and accuracy for the same settings used in Section 5, complementing the TPR and FPR presented in Tables 1, 2 and 3.
	\item Following requests by the referees, Table~\ref{stab:4} provides results on a larger simulation setting, with $p=400, n=120$, for the hubs and cliques negative structures. 
\end{enumerate}

\label{sec:sup_simulation}

\begin{table}[H]
	\centering
	\caption{Sensitivity (true positive/(true positive+false negative)=$TP/(TP+FN)$), specificity (true negative/(true negative+false positive)=$TN/(TN+FP)$), precision ($TP/(TP+FP)$), and accuracy ($(TP+TN)/(TP+TN+FP+FN)$) of precision matrix estimates over 50 data sets generated by multivariate normal distributions with precision matrix $\Omega_0$, where $p=100$ and $n=50$. The precision matrix is estimated by frequentist graphical lasso with penalized diagonal elements (GL1), frequentist graphical lasso with unpenalized diagonal elements (GL2), graphical SCAD (GSCAD), Bayesian graphical lasso (BGL), and graphical horseshoe (GHS). The best performer in each row is shown in bold.}
	
	\label{stab:1}
	\begin{footnotesize}
		\noindent\makebox[\textwidth]{%
			\begin{tabular}{l|rrrrr|rrrrr|}
				\toprule
				& \multicolumn{5}{c|}{Random} & \multicolumn{5}{c|}{Hubs} \\
				nonzero pairs & \multicolumn{5}{c|}{35/4950} & \multicolumn{5}{c|}{90/4950} \\
				nonzero elements & \multicolumn{5}{c|}{$\sim -\mathrm{Unif}(0.2,1)$} & \multicolumn{5}{c|}{0.25} \\
				$p=100, n=50$ & GL1 & GL2 & GSCAD & BGL & GHS & GL1 & GL2 & GSCAD & BGL & GHS \\
				\toprule   	
				
				SEN & .8246 & .7097 & \textbf{.9977} & .8709 & .5903 
				& .8649 & .7333 & \textbf{.9987} & .8513 & .2687 \\		
				& (.0520) & (.0620) & (.0078) & (.0470) & (.0537)
				& (.0443) & (.0751) & (.0053) & (.0378) & (.0764) \\
				
				SPE & .9053 & .9626 & .0045 & .8945 & \textbf{.9996}
				& .9081 & .9719 & .0024 & .8811 & \textbf{.9987} \\
				& (.0141) & (.0070) & (.0102) & (.0059) & (.0004)
				& (.0130) & (.0086) & (.0069) & (.0058) & (.0006)\\
				
				PREC & .0593 & .1213 & .0071 & .0556 & \textbf{.9134}
				& .1503 & .3378 & .0182 & .1172 & \textbf{.8031} \\
				& (.0073) & (.0166) & ($<$.0001) & (.0039) & (.0626)
				& (.0166) & (.0559) & ($<$.0001) & (.0057) & (.0677) \\
				ACC & .9048 & .9608 & .0116 & .8943 & \textbf{.9967}
				& .9074 & .9676 & .0205 & .8806 & \textbf{.9855} \\
				& (.0138) & (.0067) & (.0101) & (.0058) & (.0005)
				& (.0123) & (.0075) & (.0067) & (.0055) & (.0012) \\
				
				\toprule
				& \multicolumn{5}{c|}{Cliques positive} & \multicolumn{5}{c|}{Cliques negative} \\
				nonzero pairs & \multicolumn{5}{c|}{30/4950} & \multicolumn{5}{c|}{30/4950} \\
				nonzero elements & \multicolumn{5}{c|}{-0.45} & \multicolumn{5}{c|}{0.75} \\
				$p=100, n=50$ & GL1 & GL2 & GSCAD & BGL & GHS & GL1 & GL2 & GSCAD & BGL & GHS \\
				\toprule
				
				SEN & \textbf{1} & \textbf{1} & \textbf{1} & \textbf{1} & .7487 
				& .9993 & .9880 & \textbf{1} & .9993 & .9733 \\
				& (0) & (0) & (0) & (0) & (.0427)
				& (.0047) & (.0221) & (0) & (.0047) & (.0421) \\
				
				SPE & .9100 & .9745 & .0099 & .8986 & \textbf{.9997} 
				& .9078 & .9721 & .0248 & .8839 & \textbf{.9990} \\
				& (.0098) & (.0056) & (.0177) & (.0052) & (.0003)
				& (.0135) & (.0084) & (.0219) & (.0051)  & (.0005) \\
				PREC & .0641 & .1991 & .0061 & .0569 & \textbf{.9352}
				& .0632 & .1881 & .0062 & .0500 & \textbf{.8611} \\
				& (.0067) & (.0365) & (.0001) & (.0027) & (.0541) 
				& (.0090) & (.0448) & (.0001) & (.0021) & (.0615) \\
				ACC & .9106 & .9747 & .0159 & .8992 & \textbf{.9981} 
				& .9084 & .9722 & .0307 & .8846 & \textbf{.9988} \\
				& (.0097) & (.0055) & (.0176) & (.0052) & (.0004)
				& (.0135) & (.0083) & (.0218) & (.0051) & (.0005) \\
				\bottomrule
		\end{tabular}}
	\end{footnotesize}
\end{table}

\begin{table}[H]
	\centering
	\caption{Sensitivity (true positive/(true positive+false negative)=$TP/(TP+FN)$), specificity (true negative/(true negative+false positive)=$TN/(TN+FP)$), precision ($TP/(TP+FP)$), and accuracy ($(TP+TN)/(TP+TN+FP+FN)$) of precision matrix estimates over 50 data sets generated by multivariate normal distributions with precision matrix $\Omega_0$, where $p=100$ and $n=120$. The precision matrix is estimated by frequentist graphical lasso with penalized diagonal elements (GL1), frequentist graphical lasso with unpenalized diagonal elements (GL2), refitted graphical lasso (RGL), graphical SCAD (GSCAD), Bayesian graphical lasso (BGL), and graphical horseshoe (GHS). The best performer in each row is shown in bold.}
	
	\label{stab:2}
	\begin{footnotesize}
		\noindent\makebox[\textwidth]{%
			\addtolength{\tabcolsep}{-3pt}
			\begin{tabular}{l|rrrrrr|rrrrrr|}
				\toprule
				& \multicolumn{6}{c|}{Random} & \multicolumn{6}{c|}{Hubs} \\
				nonzero pairs & \multicolumn{6}{c|}{35/4950} & \multicolumn{6}{c|}{90/4950} \\
				nonzero elements & \multicolumn{6}{c|}{$\sim -\mathrm{Unif}(0.2,1)$} & \multicolumn{6}{c|}{0.25} \\
				$p=100, n=120$ & GL1 & GL2 & RGL & GSCAD & BGL & GHS & GL1 & GL2 & RGL & GSCAD & BGL & GHS \\
				\toprule   	
				SEN & .9486 & .8794 & .6497 & \textbf{.9994} & .9760 & .8149 
				& .9936 & .9844 & .8376 & \textbf{.9998} & .9938 & .8671 \\		
				& (.0316) & (.0384) & (.0658) & (.0040) & (.0233) & (.0397)
				& (.0078) & (.0154) & (.0617) & (.0016) & (.0072) & (.0396) \\
				
				SPE & .8971 & .9558 & .9891 & .0017 & .8322 & \textbf{.9995}
				& .8971 & .9569 & \textbf{.9985} & .0012 & .8128 & .9967 \\
				& (.0150) & (.0077) & (.0029) & (.0055) & (.0066) & (.0002)
				& (.0161) & (.0093) & (.0009) & (.0029) & (.0066) & (.0011) \\
				
				PREC & .0627 & .1265 & .3075 & .0071 & .0398 & \textbf{.9248} 
				& .1541 & .3044 & \textbf{.9131} & .0182 & .0896 & .8334 \\
				& (.0084) & (.0170) & (.0521) & ($<$.0001) & (.0018) & (.0392) 
				& (.0191) & (.0490) & (.0408) & ($<$.0001) & (.0029) & (.0476) \\
				
				ACC & .8975 & .9552 & .9867 & .0088 & .8332 & \textbf{.9982}
				& .8988 & .9574 & \textbf{.9955} & .0193 & .8161 & .9944 \\
				& (.0149) & (.0075) & (.0026) & (.0054) & (.0065) & (.0004)
				& (.0158) & (.0091) & (.0010) & (.0029) & (.0064) & (.0010) \\
				
				\toprule
				& \multicolumn{6}{c|}{Cliques positive} & \multicolumn{6}{c|}{Cliques negative} \\
				nonzero pairs & \multicolumn{6}{c|}{30/4950} & \multicolumn{6}{c|}{30/4950} \\
				nonzero elements & \multicolumn{6}{c|}{-0.45} & \multicolumn{6}{c|}{0.75} \\
				$p=100, n=120$ & GL1 & GL2 & RGL & GSCAD & BGL & GHS & GL1 & GL2 & RGL & GSCAD & BGL & GHS \\
				\toprule
				SEN & \textbf{1} & \textbf{1} & \textbf{1} & \textbf{1} & \textbf{1} & .9840
				& \textbf{1} & \textbf{1} & .9947 & \textbf{1} & \textbf{1} & \textbf{1} \\
				& (0) & (0) & (0) & (0) & (0) & (.0236)
				& (0) & (0) & (.0170) & (0) & (0) & (0) \\
				
				SPE & .9001 & .9713 & \textbf{.9997} & .0021 & .8420 & .9996
				& .8959 & .9587 & .9990 & .0061 & .8224 & \textbf{.9992} \\
				& (.0078) & (.0064) & (.0004) & (.0061) & (.0074) & (.0002)
				& (.0100) & (.0088) & (.0008) & (.0073) & (.0068) & (.0004) \\
				
				PREC & .0579 & .1812 & \textbf{.9574} & .0061 & .0372 & .9459
				& .0558 & .1331 & .8728 & .0061 & .0332 & \textbf{.8882} \\
				& (.0046) & (.0335) & (.0571) & ($<$.0001) & (.0017) & (.0388)
				& (.0053) & (.0241) & (.0927) & ($<$.0001) & (.0012) & (.0537) \\
				
				ACC & .9007 & .9715 & \textbf{.9997} & .0082 & .8430 & .9996
				& .8966 & .9590 & .9990 & .0121 & .8234 & \textbf{.9992} \\
				& (.0078) & (.0063) & (.0004) & (.0060) & (.0073) & (.0002) 
				& (.0099) & (.0088) & (.0008) & (.0072) & (.0067) & (.0004) \\
				\bottomrule
		\end{tabular}}
		\addtolength{\tabcolsep}{3pt}
	\end{footnotesize}
\end{table}

\begin{table}[H]
	\centering
	\caption{Sensitivity (true positive/(true positive+false negative)=$TP/(TP+FN)$), specificity (true negative/(true negative+false positive)=$TN/(TN+FP)$), precision ($TP/(TP+FP)$), and accuracy ($(TP+TN)/(TP+TN+FP+FN)$) of precision matrix estimates over 50 data sets generated by multivariate normal distributions with precision matrix $\Omega_0$, where $p=200$ and $n=120$. The precision matrix is estimated by frequentist graphical lasso with penalized diagonal elements (GL1), frequentist graphical lasso with unpenalized diagonal elements (GL2), graphical SCAD (GSCAD), Bayesian graphical lasso (BGL), and graphical horseshoe (GHS). The best performer in each row is shown in bold.}
	
	\label{stab:3}
	\begin{footnotesize}
		\noindent\makebox[\textwidth]{%
			\begin{tabular}{l|rrrrr|rrrrr|}
				\toprule
				& \multicolumn{5}{c|}{Random} & \multicolumn{5}{c|}{Hubs} \\
				nonzero pairs & \multicolumn{5}{c|}{29/19900} & \multicolumn{5}{c|}{180/19900} \\
				nonzero elements & \multicolumn{5}{c|}{$\sim -\mathrm{Unif}(0.2,1)$} & \multicolumn{5}{c|}{0.25} \\
				$p=200, n=120$ & GL1 & GL2 & GSCAD & BGL & GHS & GL1 & GL2 & GSCAD & BGL & GHS \\
				\toprule   	
				SEN & .9476 & .8393 & \textbf{1} & .9855 & .8421
				& .9911 & .9773 & \textbf{1} & .9917 & .7754 \\
				& (.0370) & (.0301) & (0) & (.0232) & (.0369)
				& (.0065) & (.0132) & (0) & (.0060) & (.0323) \\
				
				SPE & .9486 & .9841 & .0049 & .8965 & \textbf{.9999}
				& .9343 & .9743 & .0002 & .8803 & \textbf{.9989} \\
				& (.0065) & (.0021) & (.0095) & (.0031) & ($<$.0001) 
				& (.0053) & (.0064) & (.0002) & (.0027) & (.0002) \\
				
				PREC & .0265 & .0722 & .0015 & .0137 & \textbf{.9334} 
				& .1218 & .2662 & .0090 & .0703 & \textbf{.8693} \\
				& (.0031) & (.0077) & ($<$.0001) & ($<$.0001) & (.0463)
				& (.0105) & (.0467) & ($<$.0001) & (.0014) & (.0273) \\
				
				ACC & .9486 & .9838 & .0064 & .8967 & \textbf{.9997} 
				& .9348 & .9743 & .0093 & .8813 & \textbf{.9969} \\
				& (.0065) & (.0021) & (.0094) & (.0030) & ($<$.0001)
				& (.0052) & (.0063) & (.0002) & (.0027) & (.0003) \\
				
				\toprule
				& \multicolumn{5}{c|}{Cliques positive} & \multicolumn{5}{c|}{Cliques negative} \\
				nonzero pairs & \multicolumn{5}{c|}{60/19900} & \multicolumn{5}{c|}{60/19900} \\
				nonzero elements & \multicolumn{5}{c|}{-0.45} & \multicolumn{5}{c|}{0.75} \\
				$p=200, n=120$ & GL1 & GL2 & GSCAD & BGL & GHS & GL1 & GL2 & GSCAD & BGL & GHS \\
				\toprule
				
				SEN & \textbf{1} & \textbf{1} & \textbf{1} & \textbf{1} & .9633
				& \textbf{1} & \textbf{1} & \textbf{1} & \textbf{1} & \textbf{1} \\
				& (0) & (0) & (0) & (0) & (.0226)
				& (0) & (0) & (0) & (0) & (0) \\
				
				SPE & .9337 & .9828 & .0031 & .9014 & \textbf{.9998}
				& .9364 & .9771 & .0081 & .8845 & \textbf{.9996} \\
				& (.0044) & (.0027) & (.0051) & (.0030) & (.0001)
				& (.0039) & (.0039) & (.0072) & (.0027) & (.0001) \\
				
				PREC & .0438 & .1516 & .0030 & .0298 & \textbf{.9465}
				& .0455 & .1198 & .0030 & .0255 & \textbf{.8973} \\
				& (.0027) & (.0180) & ($<$.001) & (.0008) & (.0312)
				& (.0023) & (.0209) & ($<$.0001) & (.0005) & (.0367) \\
				
				ACC & .9339 & .9828 & .0061 & .9017 & \textbf{.9997}
				& .9366 & .9772 & .0111 & .8849 & \textbf{.9996} \\
				& (.0044) & (.0027) & (.0051) & (.0030) & (.0001)
				& (.0038) & (.0039) & (.0072) & (.0027) & (.0001) \\
				
				\bottomrule
		\end{tabular}}
	\end{footnotesize}
\end{table}

\begin{table}[H]
	\centering
	\caption{Mean (sd) Stein's loss, Frobenius norm,, sensitivity (true positive/(true positive+false negative)=$TP/(TP+FN)$), specificity (true negative/(true negative+false positive)=$TN/(TN+FP)$), precision ($TP/(TP+FP)$), and accuracy ($(TP+TN)/(TP+TN+FP+FN)$) of precision matrix estimates over 20 data sets generated by multivariate normal distributions with precision matrix $\Omega_0$, where $p=400$ and $n=120$. The precision matrix is estimated by frequentist graphical lasso with penalized diagonal elements (GL1), frequentist graphical lasso with unpenalized diagonal elements (GL2), graphical SCAD (GSCAD), Bayesian graphical lasso (BGL), and graphical horseshoe (GHS). The best performer in each row is shown in bold. Average CPU time is in minutes.}
	
	\label{stab:4}
	\begin{footnotesize}
		\noindent\makebox[\textwidth]{%
			\begin{tabular}{l|rrrrr|rrrrr|}
				\toprule
				& \multicolumn{5}{c|}{Hubs} & \multicolumn{5}{c|}{Cliques negative} \\
				nonzero pairs & \multicolumn{5}{c|}{360/79800} & \multicolumn{5}{c|}{120/79800} \\
				nonzero elements & \multicolumn{5}{c|}{0.25} & \multicolumn{5}{c|}{0.75} \\
				$p=400, n=120$ & GL1 & GL2 & GSCAD & BGL & GHS & GL1 & GL2 & GSCAD & BGL & GHS \\
				\toprule
				
				Stein's loss & 29.27 & 35.53 & 28.75 & 333.98 & \textbf{26.92}
				& 33.90 & 42.91 & 33.71 & 352.22 & \textbf{8.36} \\
				& (0.90) & (0.55) & (1.00) & (2.34) & (0.97)
				& (0.74) & (1.04) & (0.73) & (3.03) & (0.62) \\
				
				F norm & 7.07 & 8.05 & 6.96 & 12.09 & \textbf{5.60}
				& 11.30 & 12.80 & 11.28 & 12.92 & \textbf{4.00} \\
				& (0.19) & (0.05) & (0.21) & (0.19) & (0.11)
				& (0.23) & (0.10) & (0.22) & (0.11) & (0.16) \\
				
				SEN & .9826 & .9569 & \textbf{1} & .9871 & .6844
				& \textbf{1} & \textbf{1} & \textbf{1} & \textbf{1} & \textbf{1} \\
				& (.0092) & (.0130) & (0) & (.0063) & (.0302)
				& (0) & (0) & (0) & (0) & (0) \\
				
				SPE & .9651 & .9882 & .0001 & .9378 & \textbf{.9996}
				& .9580 & .9884 & .0102 & .9359 & \textbf{.9998} \\
				& (.0065) & (.0006) & ($<$.0001) & (.0011) & ($<$.0001)
				& (.0059) & (.0015) & (.0119) & (.0010) & ($<$.0001) \\
				
				PREC & .1157 & .2690 & .0045 & .0671 & \textbf{.8884}
				& .0353 & .1163 & .0015 & .0230 & \textbf{.9052} \\
				& (.0166) & (.0111) & ($<$.0001) & (.0013) & (.0221)
				& (.0052) & (.0103) & ($<$.0001) & (.0003) & (.0228) \\
				
				ACC & .9652 & .9880 & .0046 & .9380 & \textbf{.9982}
				& .9581 & .9885 & .0117 & .9360 & \textbf{.9998} \\
				& (.0065) & (.0006) & ($<$.0001) & (.0011) & (.0001)
				& (.0059) & (.0015) & (.0119) & (.0010) & ($<$.0001) \\
				
				Avg CPU time & 0.22 & 0.25 & 2.03e+3 & 5.48e+3 & 6.63e+3
				& 0.30 & 0.33 & 2.01e+3 & 5.81e+3 & 5.52e+3 \\
				
				\bottomrule
		\end{tabular}}
	\end{footnotesize}
\end{table}

\end{document}